# Grain Boundary Segregation Spectra from a Generalized Machine-learning Potential

Nutth Tuchinda[a], Christopher A. Schuh[b, a*]

[a]Department of Materials Science and Engineering, Massachusetts Institute of Technology, 77 Massachusetts Avenue, Cambridge, MA, 02139, USA

[b]Department of Materials Science and Engineering, Northwestern University, 2145 Sheridan Road, Evanston, IL 60208, USA

*Correspondence to schuh@northwestern.edu

## Graphical Abstract

**Polycrystalline Model**

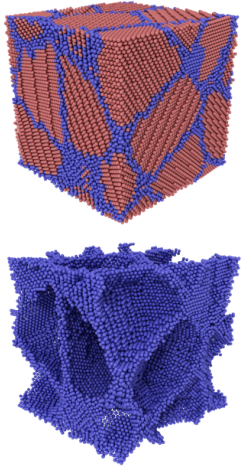

**Accelerated GB Spectra**

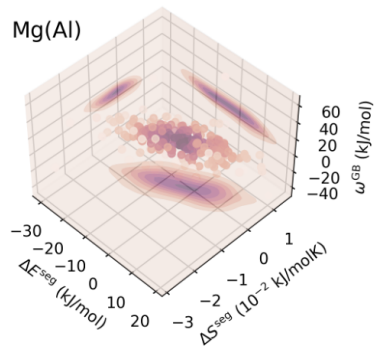

**Isotherm for GB concentration**

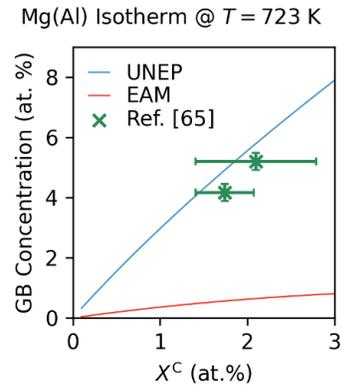

## Abstract

Modeling solute segregation to grain boundaries at near first-principles accuracy is a daunting task, particularly at finite concentrations and temperatures that require accurate assessments of solute-solute interactions and excess vibrational entropy of segregation that are computationally intensive. Here, we apply a generalized machine learning potential for 16 elements, including Ag, Al, Au, Cr, Cu, Mg, Mo, Ni, Pb, Pd, Pt, Ta, Ti, V, W and Zr, to provide a self-consistent spectral database for all of these energetic components in of 240 binary alloy polycrystals. The segregation spectra of Al-based alloys are validated against past quantum-accurate simulations and show improved predictive ability with some existing atom probe tomography experimental data.

The spectral model for grain boundary GB segregation has been recently developed as a rigorous thermodynamic framework for treating segregation in polycrystals [1,2]. At its core, the spectral model considers a broad spectrum of GB atomic environments, each of which has its own set of thermodynamic quantities including segregation enthalpy [2–4], excess entropy [5–9], and solute-solute interaction parameters [10–13]. The field has recently focused on tabulating those thermodynamic spectra for many alloys, and the vast complexity of the problem is substantially traversed through the use of data science or machine learning (ML) methods [14–17]. The resulting learned GB thermodynamic spectra form the basis for myriad downstream applications, from predicting GB segregation states to alloy design [18–21].

Despite the progress in developing the GB spectral model, to date there have been a limited number of databases for alloy design developed, in large part due to a lack of accurate interatomic potentials that can describe defect chemistry across multiple transition metal elements [18–20,22].Developing an accurate potential for the specific case of GB environments is a challenge for any individual system [23–25], and the quality of the potentials used to general GB segregation spectra continues to be a major concern for progress in this area [18,26,27]. Although there are promising approaches such as hybrid quantum mechanical-molecular mechanic (QM-MM) methods [22,28] to treat the 0 K enthalpy part of the GB segregation problem, the development of full spectra of all GB properties (including site-wise entropies relevant to high temperatures, and solute-solute interaction parameters) rely on potentials.

A key advantage of the ML models developed for GB segregation spectra is that these are, in fact, meta-models: they take interatomic potentials as inputs, and they return quantitative models of GB thermodynamics. Therefore, as newer and more accurate potentials arise, the quantitative power of segregation modeling rises. In particular, the emergence of ML methods in interatomic potential development promises the rapid expansion of accurate potentials for many alloys, which can then be used to generate accurate GB thermodynamic data. In this letter, we apply a generalized interatomic potential for 16-element alloy combinations, to generate a database that is consistent across 240 binary alloy pairs using accelerated spectral models [18–20]. The database includes a full tri-variate distribution of per-site dilute segregation energies, solute-solute interaction energies, and vibrational excess entropies for defect engineering in metallurgy.

The workflow is summarized in Fig. 1. We evaluate grain boundary segregation spectra using accelerated frameworks similar to Refs. [18–20]. We construct polycrystals with 10 grains of size 12×12×12 nm using atomsk [29], followed by zero-pressure annealing at $0.3T_{melt}$ [30] and quenching to 0 K at a cooling rate of 3 K/ps in Fig. 1a followed by a FIRE relaxation [31,32] with a force tolerance of $10^{-3}$ eV/Å. We use a generalized machine learning interatomic potential called the "unified neuroevolution potential" (UNEP, nep-44-30), which is used to describe all 240 binary alloy combinations consisting of Ag, Al, Au, Cr, Cu, Mg, Mo, Ni, Pb, Pd, Pt, Ta, Ti, V, W and Zr [30]. The molecular dynamics in this work is conducted by GPUMD and LAMMPS software packages [30,33–39], and analysis is done via Python and related software packages [14,40–46]. We use adaptive common neighbor analysis implemented in OVITO [47,48] to identify GB sites (and exclude triple junction sites), and quantify local atomic environment using smooth overlap of atomic position (SOAP [49–52]) descriptor with $n^{max} = 6$, $l^{max} = 12$, $\sigma = 1$ and $r^{cut} = 6$ Å, which are then contracted to 10-element SOAP vectors using principal component analysis [53–55]. 250 grain boundary sites selected by K-means clustering [56,57] as demonstrated in Fig. 1b are used to rapidly sample the full configuration space.

We show an example output for Ag(Al) in Fig. 1c with the full trivariate spectrum of segregation parameters for each site represented (enthalpy, excess entropy, interactions). Such distributions can be described via fitted functions such as:

$$P = \frac{1}{2\pi\sqrt{|\Sigma|}} \exp\left[-\frac{1}{2}(\boldsymbol{x} - \boldsymbol{\mu})^T \Sigma^{-1}(\boldsymbol{x} - \boldsymbol{\mu})\right] \quad (1)$$

with the probability density $P$ described by a mean ($\boldsymbol{\mu}$) and covariance ($\boldsymbol{\Sigma}$) matrix for all three components calculated in this work (see supplemental material for full data). Details on how we compute each of these thermodynamic parameters are provided in what follows. For each site $i$, we calculate dilute segregation energy ($\Delta E_i^{seg}$) by computing the energy E of four system configurations as:

$$\Delta E_i^{seg} = (E_{GB,i}^{solute} - E^{pure}) - (E_{bulk}^{solute} - E^{pure}) \quad (2)$$

where the superscripts denote whether there is a solute atom occupying the site or not, and the subscript denotes the site type (GB or bulk). $E^{pure}$ is the potential energy of the pure solvent polycrystalline system. All the configurations are relaxed using FIRE minimization with a force tolerance of $10^{-3}$ eV/Å. The dilute segregation energy spectrum for a given binary alloy is described with a skew-normal distribution [58,59] as demonstrated in Fig. 1c.:

$$P(\Delta E^{seg}) = \frac{1}{\sqrt{2\pi}\sigma} \exp\left[-\frac{(\Delta E^{seg} - \mu)^2}{2\sigma^2}\right] \text{erfc}\left[\frac{\alpha(\Delta E^{seg} - \mu)}{\sqrt{2}\sigma}\right] \quad (3)$$

where the three parameters $\alpha$, $\mu$ and $\sigma$ describes the skewness, mean and width of the distribution respectively. The skew-normal distribution parameters are calculated using the algorithm described in Ref. [2].These dilute segregation energies capture the enthalpic portion of the total segregation energy, and offer a convenient opportunity to validate the generalized ML potentials against quantum-based methods. A database of such QM-MM is available for GBs of Al binary alloys [22], providing an overlap of 15 systems with the present work for validation. Fig. 2a shows a few segregation spectra for such alloys, which show good agreement with UNEP potential in general—considerably better agreement than the agreement found with many embedded atom method (EAM) potentials [22]. The degree of matching between these methods is best analyzed by considering their thermodynamic predictive capability. The manner in which such spectra convert to local GB enrichment predictions is via a segregation isotherm, which is an integral over the spectrum occupation:

$$\overline{X}^{GB} = \int P \left[ 1 + \frac{1-X^C}{X^C} \exp\left(\frac{\Delta E^{seg}}{k_B T}\right) \right]^{-1} d\Delta E^{seg} \qquad (4)$$

where $k_B$ and $T$ are the Boltzmann constant and temperature respectively. The segregation isotherms calculated from the QM-MM database [22] vis-à-vis the present UNEP [30,60,61] are compared Fig. 2b for Al(Ni, Au, Cr and Ti) at 700 K and $f^{GB} = 0.15$ (finite grain size correction at d~20 nm [2,62]). Using a single composition of 10 at.% as a benchmark, we compare the prediction for all 16 overlapping alloys in Fig. 2c. The level of agreement is extremely satisfactory for most practical purposes [30,63].

With the generalized ML potentials providing accuracy comparable to quantum mechanical methods, we can proceed to quickly evaluate GB segregation spectra for other alloys as well. We show as an example 15 Mg-based spectra in Fig. 3, and the remaining 225 systems in the supplemental material. These spectra should help provide insight on experimental observations of GB segregation. For instance, experiments have found that Mg(Ag) segregates [64], which conforms with the negative tail for Mg(Ag) in Fig. 3. Mg(Al) shows a very strong segregation trend here with the negative tail extending more than near −50 kJ/mol. This agrees directionally with the observation that this system segregates [65,66], but experimentally such segregation is slight. As we shall see below, this is because of the critical importance of secondary segregation quantities, namely, solute-solute interaction energetics [13,67], and excess segregation entropy [7,8].

Turning first to solute-solute interactions of nearest neighbor atoms, we perform bond-wise calculations of the site-average interaction energy, $\omega_i^{GB}$, using the permutation method described in Ref. [10]:

$$\omega_i^{GB} = \frac{1}{2} \sum_j \left( E_{i-j}^{AA} + E_{i-j}^{BB} \right) - \left( E_{i-j}^{BA} + E_{i-j}^{AB} \right) \qquad (5)$$

where $E$ is the relaxed system energy with A and B dictating whether the solvent 'A' or solute 'B' occupy site $i$ (center site) and nearest neighbor site $j$ respectively. The factor ½ is needed to avoid double counting the bonds [19]. By the definition used in this paper, positive values indicate repulsive interactions between solute atoms, while negative values imply a tendency for clustering. The ML model described in Ref. [19,68] is used to rapidly compute the interaction spectra for many alloys.

For the vibrational excess entropy, we perform calculations with the LAMMPS software package using the dynamical matrix command [69,70], which allows us to estimate vibrational excess entropies from the harmonic vibrational frequencies $v_m$ [6,8,71–73]:

$$F_{\text{vib}} = k_B T \sum_m^{3N} \ln\left[2\sinh\left(\frac{h\nu_m}{2k_B T}\right)\right] \tag{6}$$

With $h$ as the Planck's constant. Here we apply a modified harmonic method to evaluate all $\nu$ near a GB site as described in Refs. [8,20] with the full and local harmonic cutoff of 13 and 16 Å respectively. The site-wise excess entropy is computed with an analogous definition to the energy:

$$\Delta S_i^{\text{seg}} = -\frac{1}{T}\left((F_{\text{vib,solvent}}^{\text{GB},i} - F_{\text{vib,pure}}^{\text{GB},i}) - (F_{\text{vib,solute}}^{\text{bulk}} - F_{\text{vib,pure}}^{\text{bulk}})\right) \tag{7}$$

with the vibrational free energies $F_{\text{vib}}$ calculated from (5) at a GB or bulk site, with either solvent or solute atoms occupying the site. The ML model described in Ref. [20] is also used to rapidly compute the excess entropy spectra for many alloys.

The outcome of these computations is a complete set of spectra for all three segregation energies—the enthalpy, excess entropy, and interactions, with a full form that is shown in Fig. 1c: the data comprise a three dimensional, trivariate probability density distribution for each alloy. In practice, however, a full trivariate distribution is not typically needed, and two correlations can be used to simplify the amount of data needed to describe an alloy. First, the vibrational excess entropies are correlated to the site enthalpies via a linear relationship [7–9,74]:

$$\Delta S_i^{\text{seg}} = \chi \Delta E_i^{\text{seg}} + \Delta S_0^{\text{solute,GB}} \tag{8}$$

with $\chi$ and $\Delta S_0^{\text{solute,GB}}$ as the slope and intercept of the linear relationship. An example of such a "enthalpy-entropy compensation" effect is shown for Al(Ag) in Fig. 4a. With this simplification, entropy effects are well-captured with two additional parameters for any given alloy. Second, the solute-solute interaction parameters are also correlated to the site enthalpies as:

$$\omega_i^{\text{GB}} = \eta \Delta E_i^{\text{seg}} + \omega_0^{\text{GB}} \tag{9}$$

With slope $\eta$ and intercept $\omega_0^{\text{GB}}$ providing a simple means of capturing solute interactions in just two parameters for any given alloy. In some alloys, there is no strong correlation [10], so the average $\omega_0^{\text{GB}}$ often suffices for approximating solute-solute interactions for use with the isotherm, as $\eta$ becomes negligible. Moreover, we can include bulk solute-solute interactions for the cases of high bulk concentrations via a similar energetic gain of $-X^C \omega^C$ where $\omega^C$ can be evaluated similarly to $\omega_i^{\text{GB}}$ for a crystalline bulk site [19,75] (again noting that negative interaction coefficients translate to attractive interactions).

With these simplifications, GB segregation in any given alloy is fully characterized by seven parameters—three for the enthalpy distribution, two for the entropy, and two for the interactions. Upon integrating Eq. (7) and (8) into (3), The final form of the isotherm then becomes:

$$\bar{X}^{\text{GB}} = \int P \left[ 1 + \frac{1 - X^C}{X^C} \right. $$
$$\left. \exp\left( \frac{\Delta E^{\text{seg}} - T\left(\chi \Delta E^{\text{seg}} + \Delta S_0^{\text{solute,GB}}\right) + \bar{X}^{\text{GB}}(\eta \Delta E^{\text{seg}} + \omega_0^{\text{GB}}) - X^C \omega^C}{k_{\text{B}} T} \right) \right]^{-1} d\Delta E^{\text{seg}} \quad (10)$$

Our ML models provide these spectra and the seven distribution fitting parameters that can be used with Eq. (9) to predict GB segregation, on average, for polycrystals, at virtually all temperatures and compositions. All 240 calculated spectra are provided in the supplemental material. While prior work has provided collections of individual enthalpy [18], or entropy [20], or interaction distributions [19], the current compilation is the first to self-consistently provide all three dimensions together for a large set of alloys using a single consistent interatomic potential. The only omissions from the full set that can be addressed with the current UNEP potentials are several Pb-based systems that give negative eigenvalues calculation of the bulk site with a solute atom. There are also several BCC solute species that exhibit very strong attractive interactions and may require further validation using a first principles method as well; these are denoted in the atlas provided in the supplemental material.

There are several limitations in the models in this work. For example, these models do not contemplate solute ordering in the GBs, which could invalidate the assumption of random-mixing that underlies the isotherms used here. We also therefore have only treated the isoconfigurational GB state here, and do not contemplate changes in the GB structure, complexions, etc. [76–78]. Moreover, GB concentrations depend on the definition of the thickness or interaction volume of the characterization methodology used [68,79] (which could be larger for experiments vis-à-vis our atomistic model (which is of order 1 nm), resulting in overestimation of $X^{\text{GB}}$ from the isotherms in Fig. 5). Nonetheless, these spectra provide new opportunities to interpret and understand GB segregation in cases where thermodynamic data have been lacking, incomplete, or inaccurate. For illustration, we show in Fig. 5 two example systems: Pt(Ni) and Mg(Al) alloys [65,80], both of which have been studied with atom probe tomography to directly quantify solute content at grain boundaries. The first three columns of Fig. 5 are the dilute segregation energy, solute-solute interactions and vibrational entropy spectra for three alloys respectively. Example isotherms are computed based on these spectra in the right-hand column, Fig. 5d and 5h. In each panel of Fig. 5, we compare the present UNEP approach to previously published EAM potentials. Clearly, the UNEP results differ significantly from those of EAM potentials, which are not typically fitted to GB properties and can therefore present significant inaccuracies.

First, we consider the Pt(Ni) system studied by Kuo et al. [80] at 850 K. Our comparison here is a generalized EAM potential from Ref. [81], which also offers a general mixing model for the 16 elements shown here. The EAM spectrum in Fig. 5a is in agreement with the accelerated model previously published with the same potential [18], and importantly, it predicts GB segregation to be very minor in this system. That prediction is considerably misaligned with experimental data, as seen in Fig. 5d, where the red EAM prediction falls far from the experimental data point. In contrast, the present UNEP potential shows a much stronger segregation spectral tail in Fig. 5a, roughly double the energetic benefit at the most segregating sites. This contrast significantly inflates the expected segregation, bringing it into much closer accord with the APT data.

Next, we turn our attention to Mg(Al) annealed at 723K in Ref. [65] and showing a segregation factor of ~ 2 vis-à-vis the bulk concentrations of ~ 2 at %. The available EAM potential predicts largely anti-segregating site enthalpies in Fig. 5e (also in agreement with the spectrum in Ref. [18] calculated from the same potential). In contrast, the present UNEP potentials capture a large segregating tail, and also show much more significant site entropies (Fig. 5f) and significantly different interactions (Fig. 5g) as well. As it turns out, all of these differences matter significantly. As seen in Fig. 5h, the EAM model dramatically underpredicts the segregation. And interestingly, if we apply only the UNEP segregation energy spectrum (dashed line in Fig. 5h), we overpredict the GB content by nearly half. It is when all of the secondary segregation quantities are included on top of the enthalpy alone that the more accurate UNEP model matches the degree of segregation produced by the experiments. In this case, the entropy is the dominant factor, as the strong site entropies in Fig. 5f oppose segregation and lower the isotherm prediction.

These two examples illustrate the strength of the accelerated GB spectral meta-model: it is easily applicable to any model potentials and can quickly evaluate and compare GB thermodynamic spectra at a fraction of cost vis-à-vis MDMC simulations. Fig. 5 therefore encourages future application of the accelerated spectral models in GB segregation engineering. The atlas of GB spectra for 240 binary alloy systems in the supplemental materials provide better spectral accuracy than classical potentials due to the capability of the machine learning potential. The framework shown here is generalized and should be applicable to any future machine learning potentials beyond 16 metallic elements in this work once they become available. We look forward to more development in this direction.

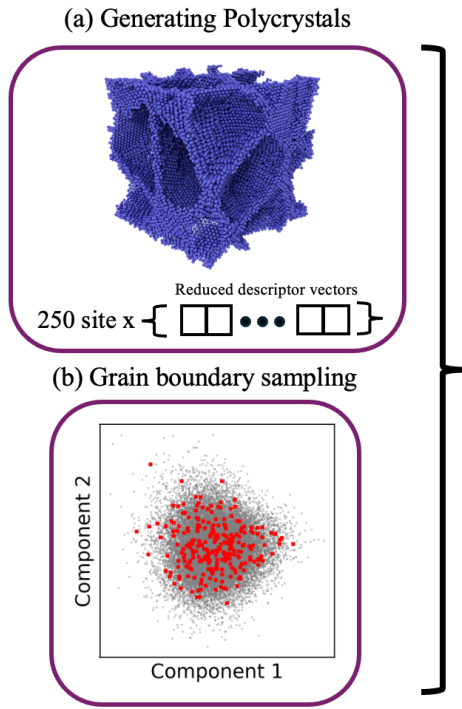
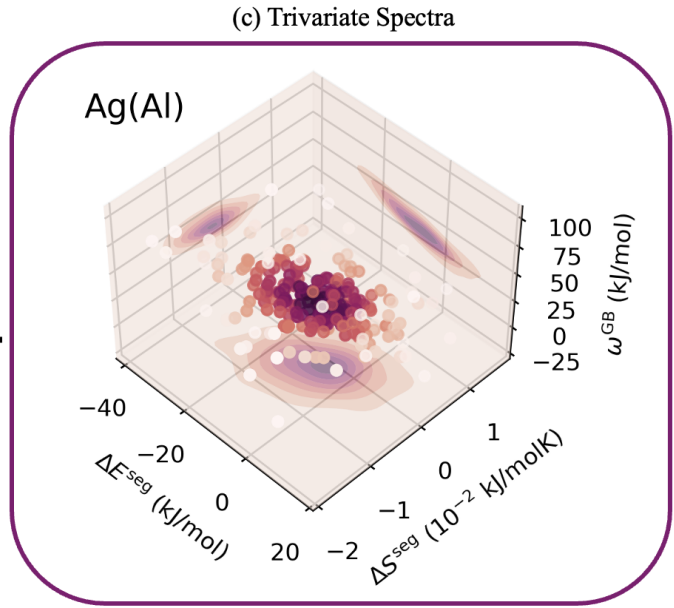

**Fig. 1** Accelerated framework for spectral grain boundary models. 250 grain boundary sites (a) are sampled from local environment descriptors (b) to assess dilute site segregation energy, site excess vibrational segregation entropy and pair-wise interactions of solute in grain boundaries (c).

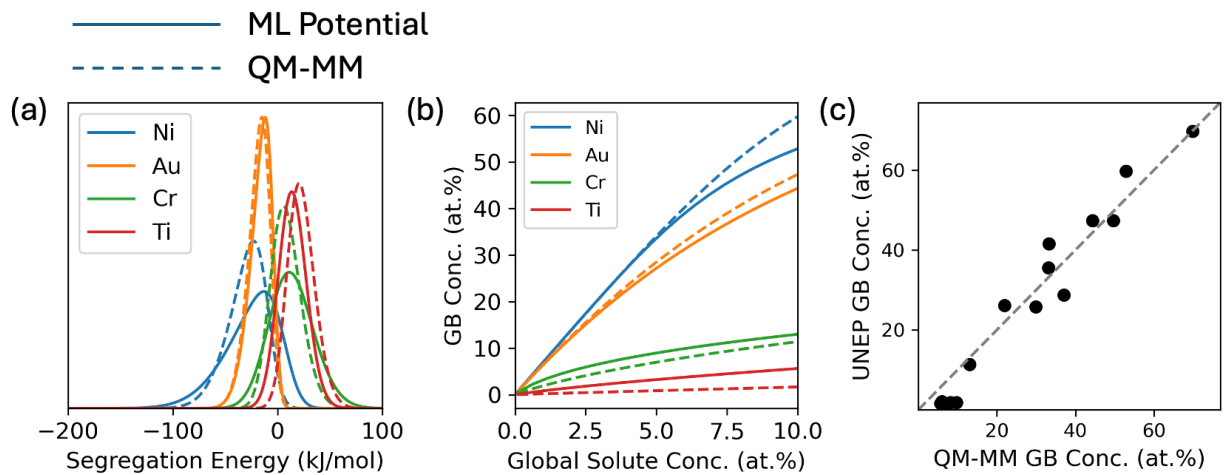

**Fig. 2** (a) Comparison of quantum accurate spectra from Ref. [22] with the spectra derived from this work via the machine learning potential from Ref. [30]. The isotherms for Al(Ni, Au, Cr and Ti) are also calculated in (b) at $T = 700$ K and $d = 20$ nm. A parity plot for predicted grain boundary concentration at $X^{tot} = 10$ at.% is also plotted in (c) using all 15 dilute segregation spectra.

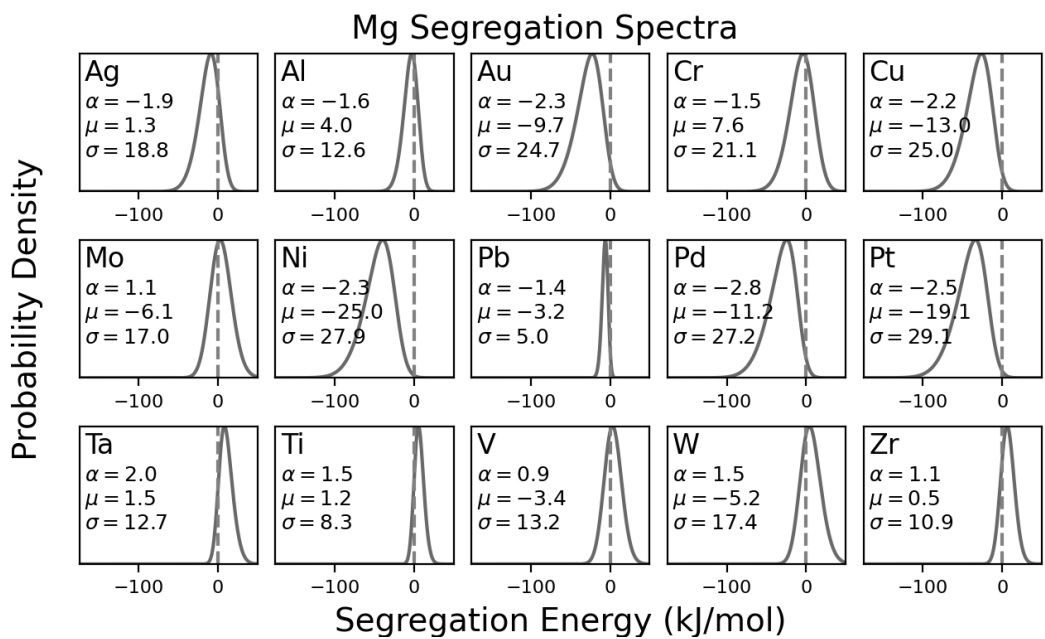

**Fig. 3** Example of all dilute segregation enthalpy spectra for Mg-based alloys calculated in this work in kJ/mol.

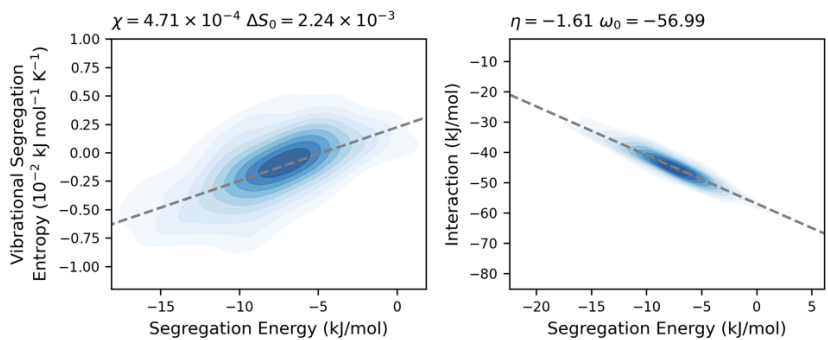

**Fig. 4** Example of correlations between the segregation energy (or enthalpy) and (a) excess vibrational entropy, (b) site solute-solute interactions for Al(Ag).

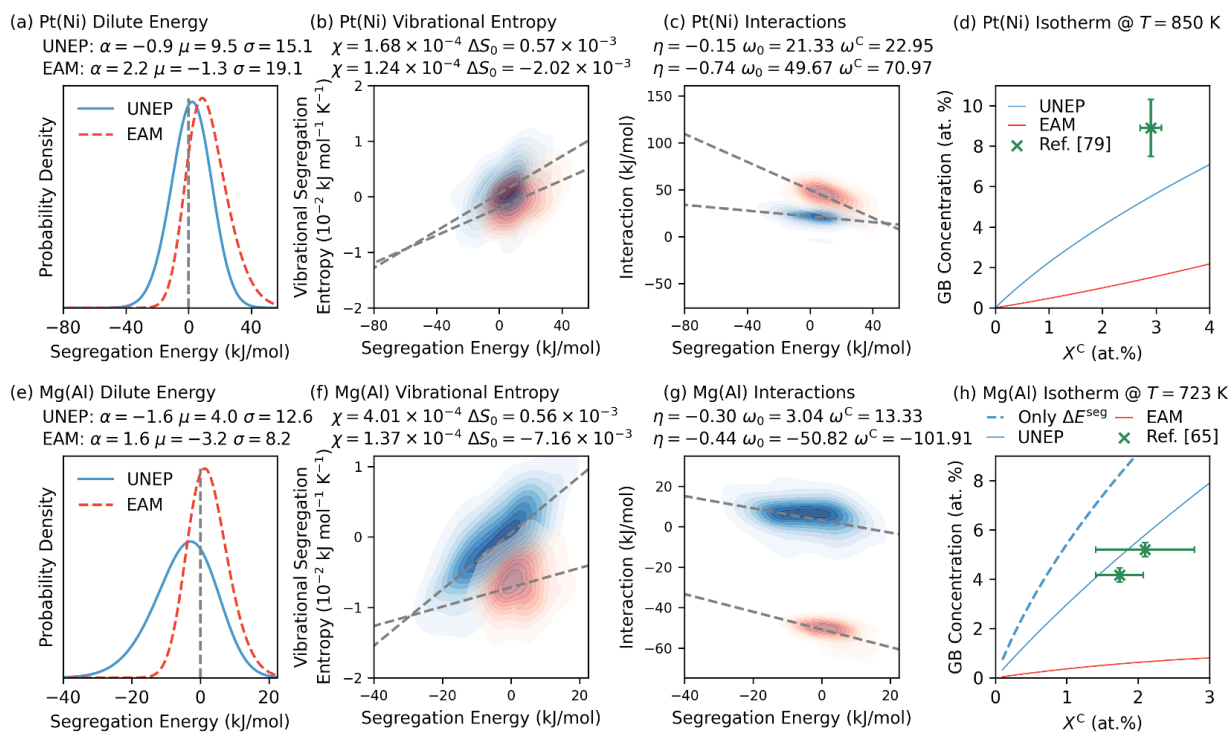

**Fig. 5** Spectral grain boundary model for Pt(Ni) (a-d) and Mg(Al) (e-h) from both generalized UNEP and EAM potentials [81]. The first three columns are the spectra for segregation energies, solute interactions and vibrational excess entropies respectively. The experimental reported values from Refs. [65,80] are noted in (d) and (h).

## Acknowledgements

This work was supported by the US Department of Energy award No. DE-SC0020180. The authors would like to acknowledge MIT ORCD, MIT supercloud [82] and MIT Satori for the HPC resources used in this work. N. Tuchinda acknowledges fruitful discussion with T. Matson at Northwestern University and C. Li at Massachusetts Institute of Technology.

## Conflict of Interest

The authors declare that they have no known competing financial interests or personal relationships that could have appeared to influence the work reported in this paper.

## Author Contributions

**Nutth Tuchinda:** Conceptualization, Data curation, Formal analysis, Investigation, Methodology, Software, Validation, Visualization, Writing – original draft, Writing – review & editing. **Christopher A. Schuh**: Conceptualization, Funding acquisition, Project administration, Resources, Supervision, Validation, Writing – original draft, Writing – review & editing

# Grain Boundary Segregation Spectra from a Generalized Machine-learning Potential

Nutth Tuchinda[a], Christopher A. Schuh[b, a*]

[a]Department of Materials Science and Engineering, Massachusetts Institute of Technology, 77 Massachusetts Avenue, Cambridge, MA, 02139, USA

[b]Department of Materials Science and Engineering, Northwestern University, 2145 Sheridan Road, Evanston, IL 60208, USA

*Correspondence to schuh@northwestern.edu

## Supplemental Material

### 1. Segregation Energy Spectra

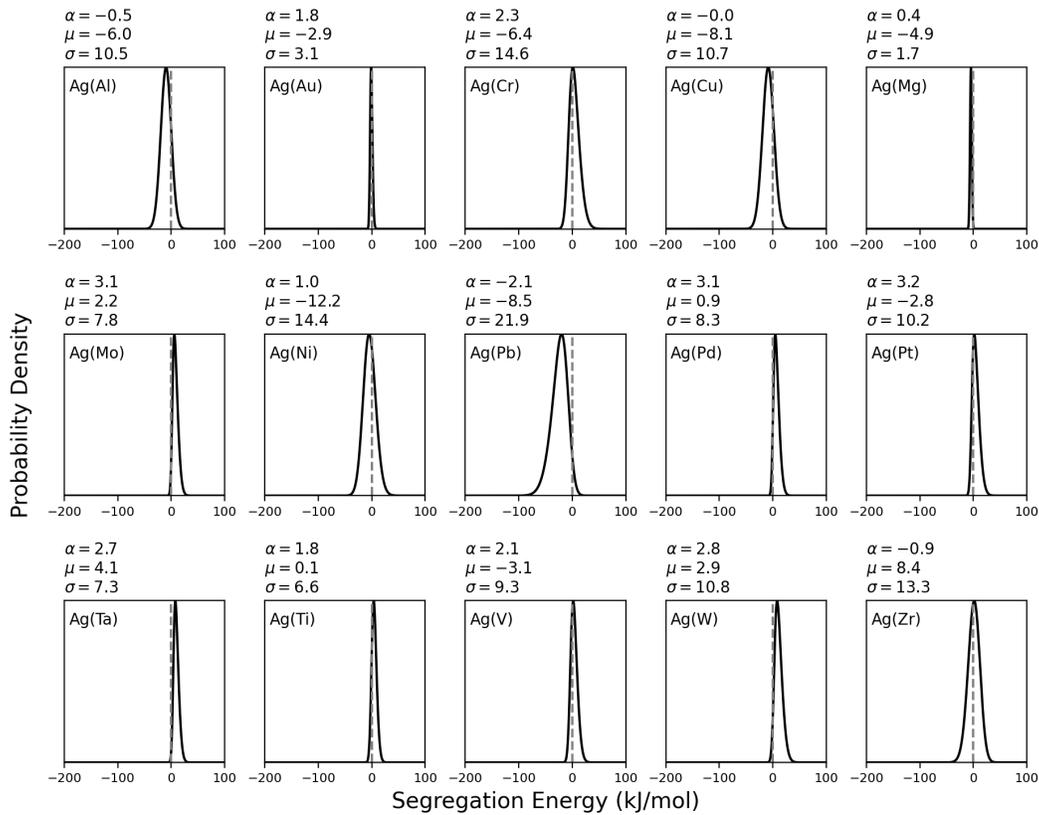

Fig. S1 Ag-based segregation energy spec in kJ/mol.

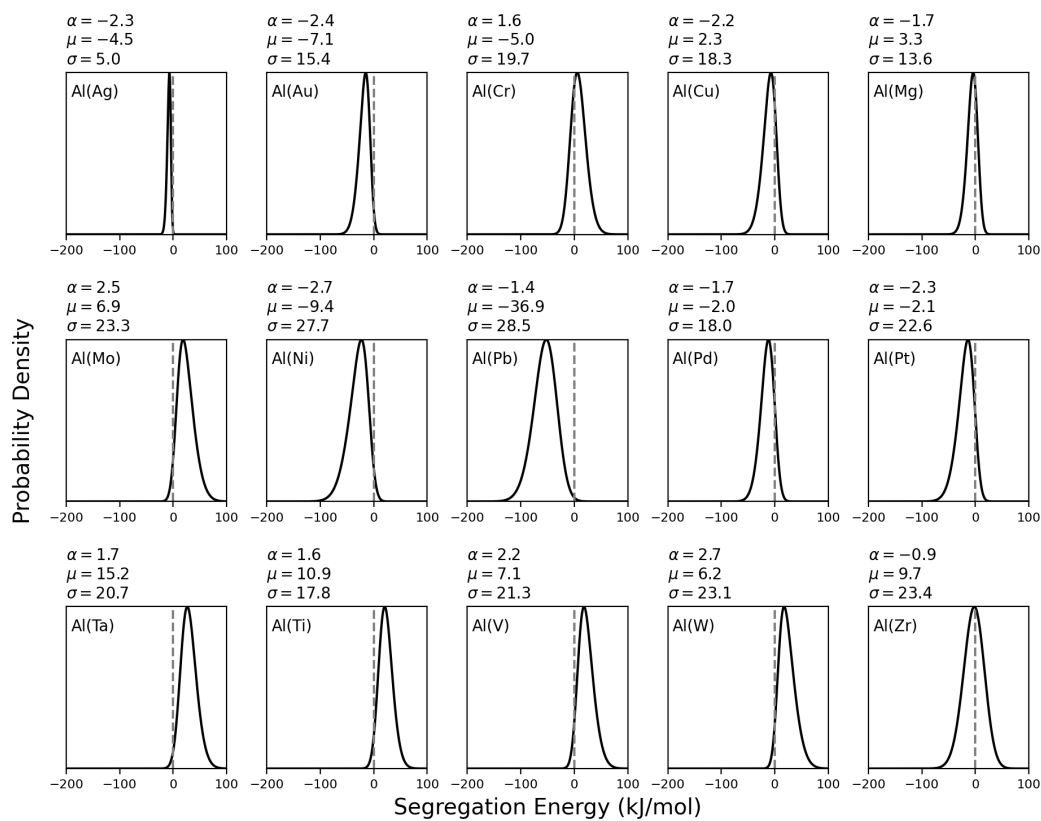

Fig. S2 Al-based segregation energy spec in kJ/mol.

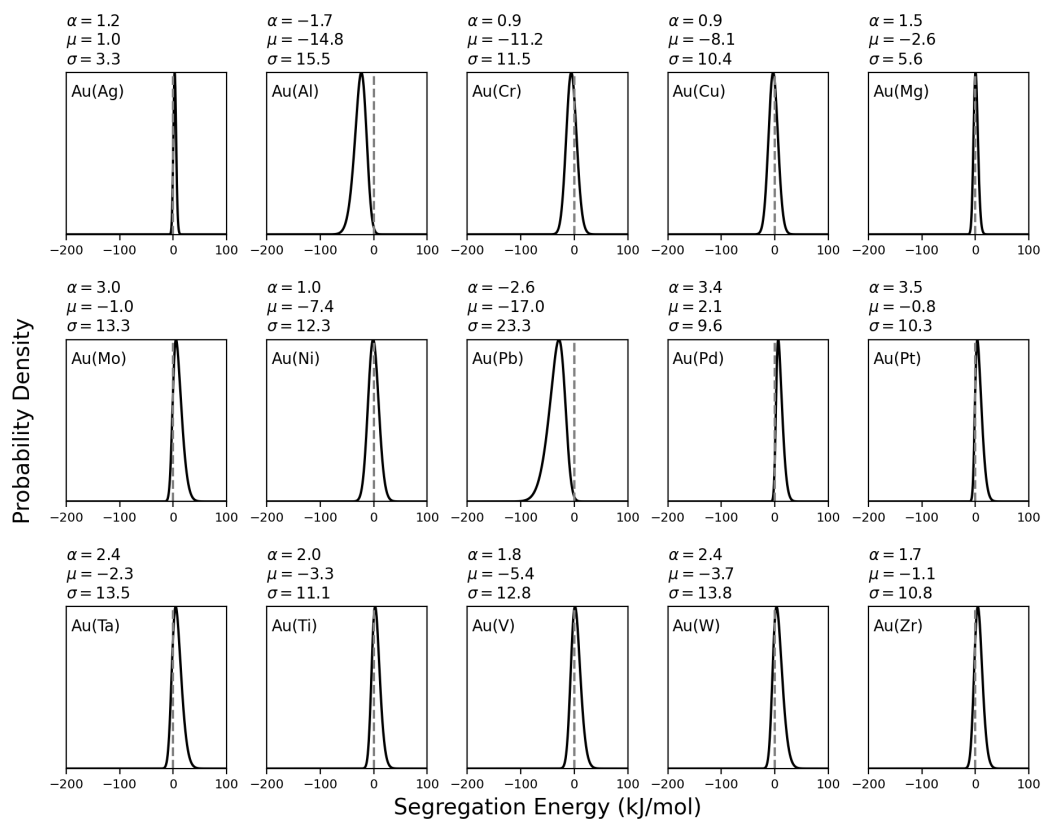

Fig. S3 Au-based segregation energy spec in kJ/mol.

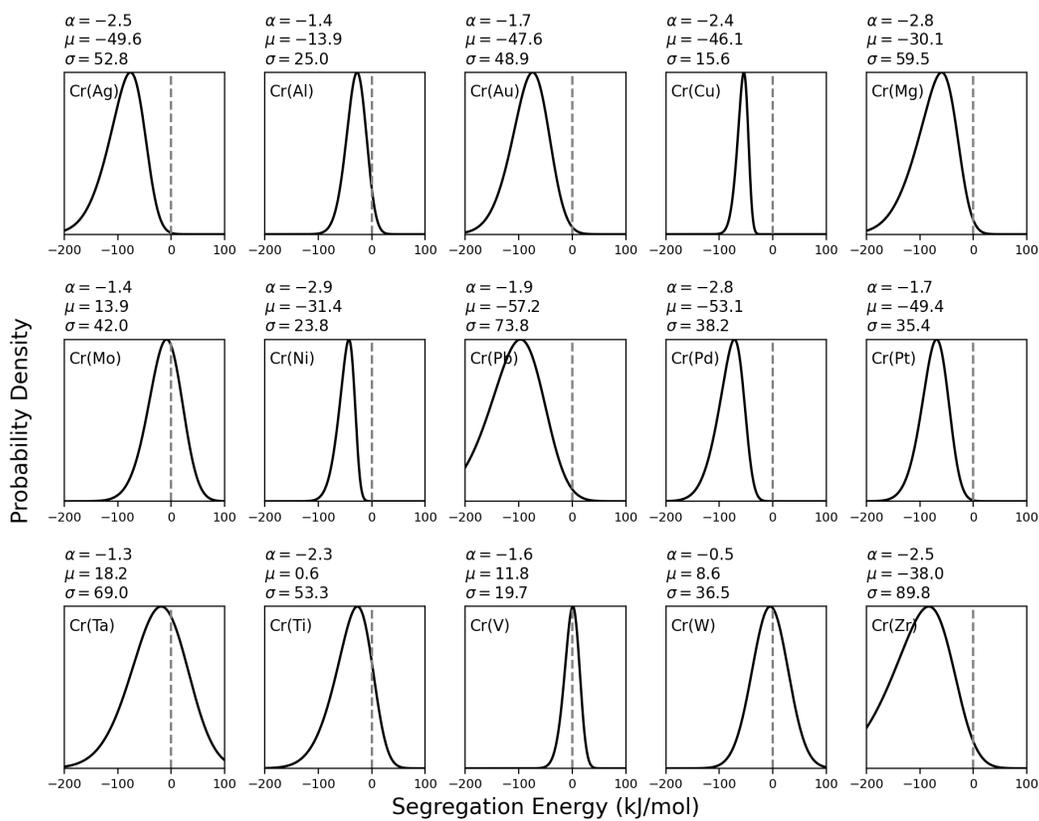

Fig. S4 Cr-based segregation energy spec in kJ/mol.

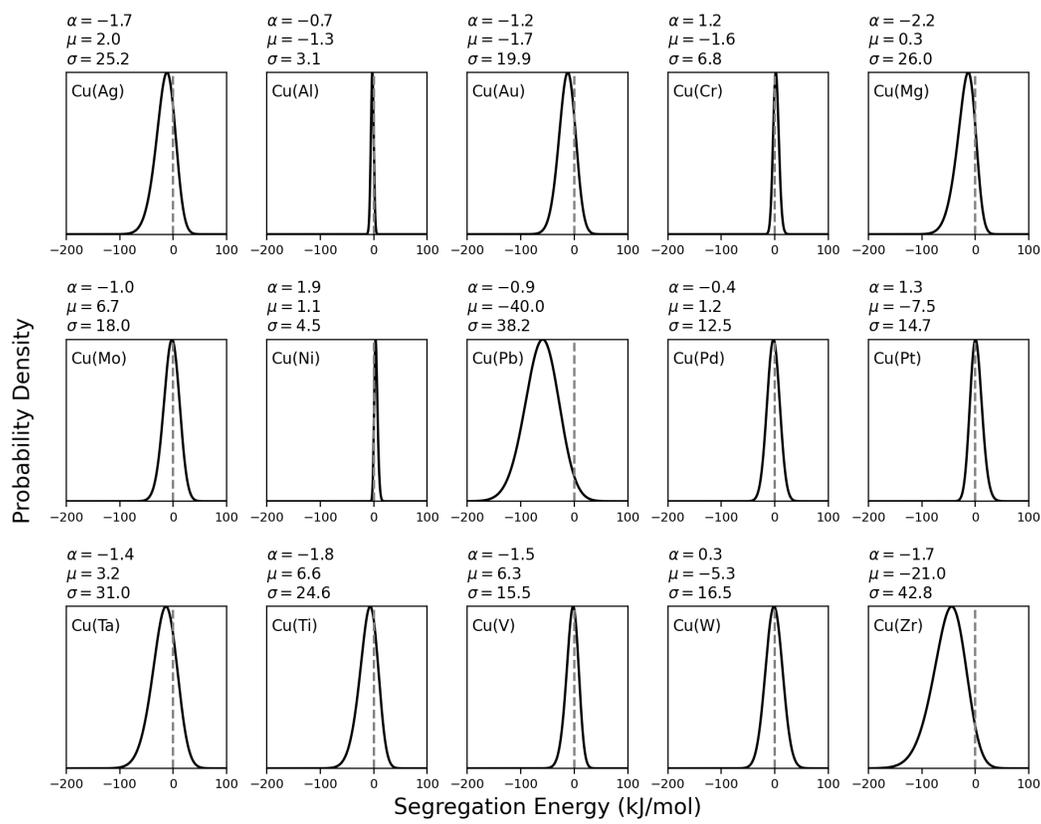

Fig. S5 Cu-based segregation energy spec in kJ/mol.

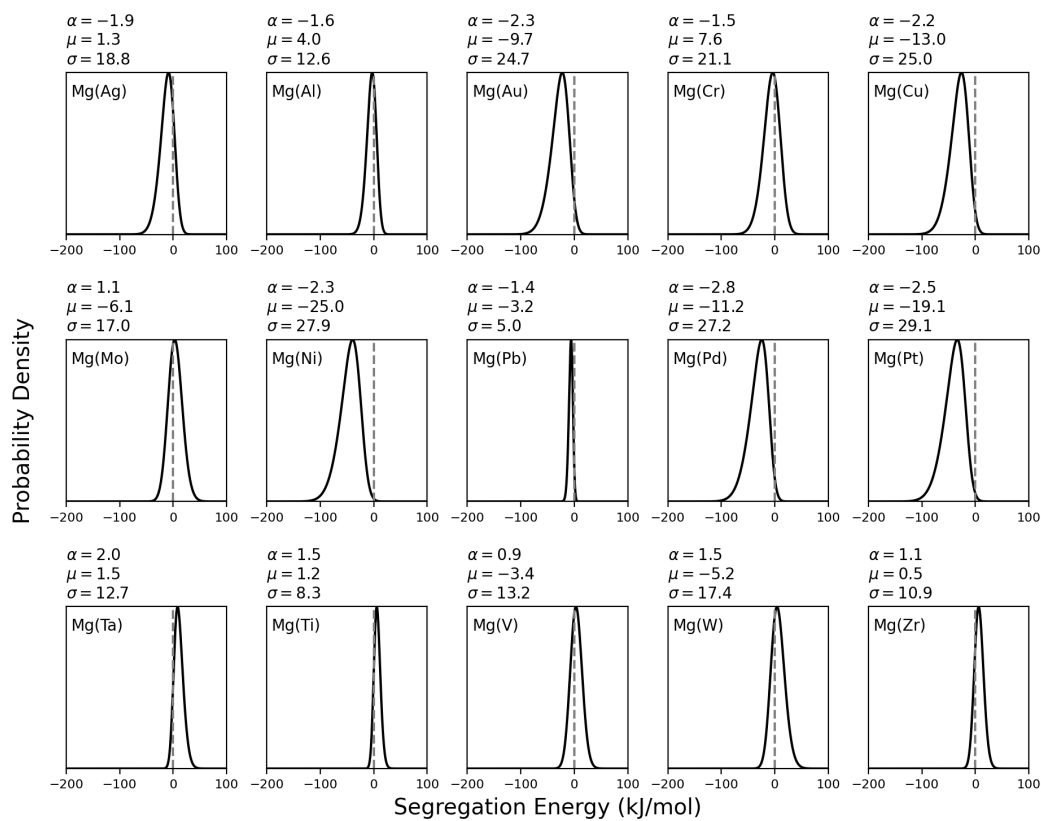

Fig S6 Mg-based segregation energy spec in kJ/mol.

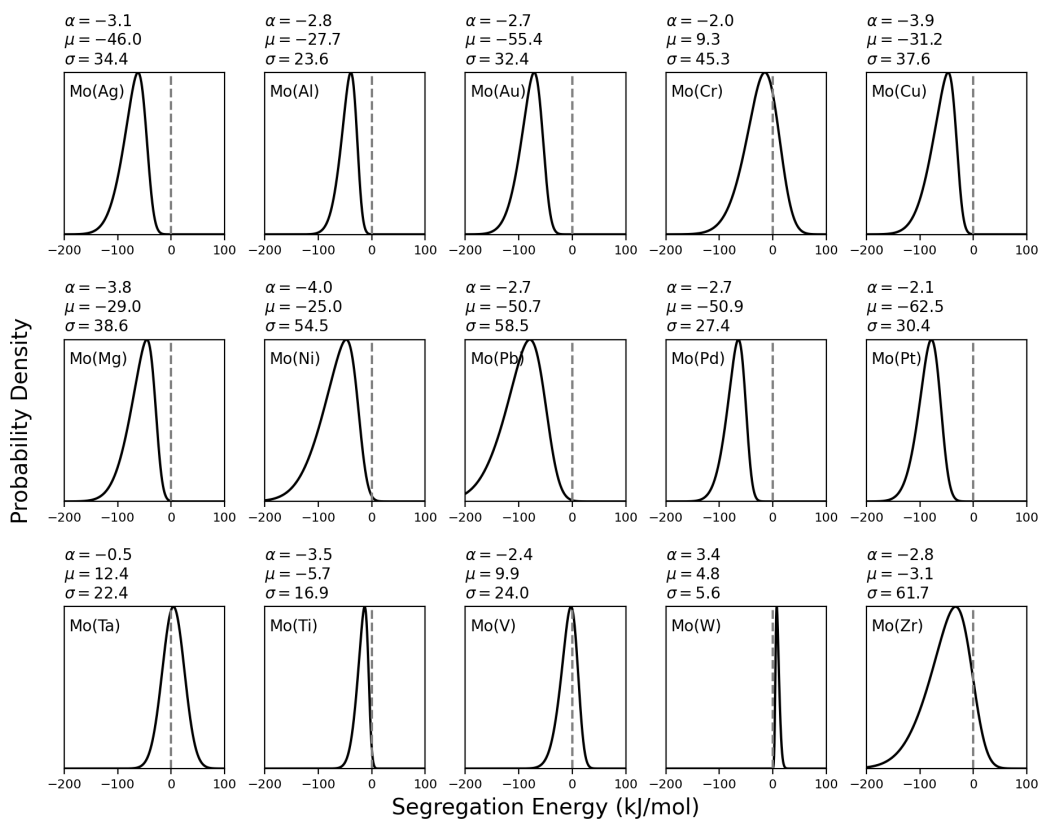

Fig. S7 Mo-based segregation energy spec in kJ/mol.

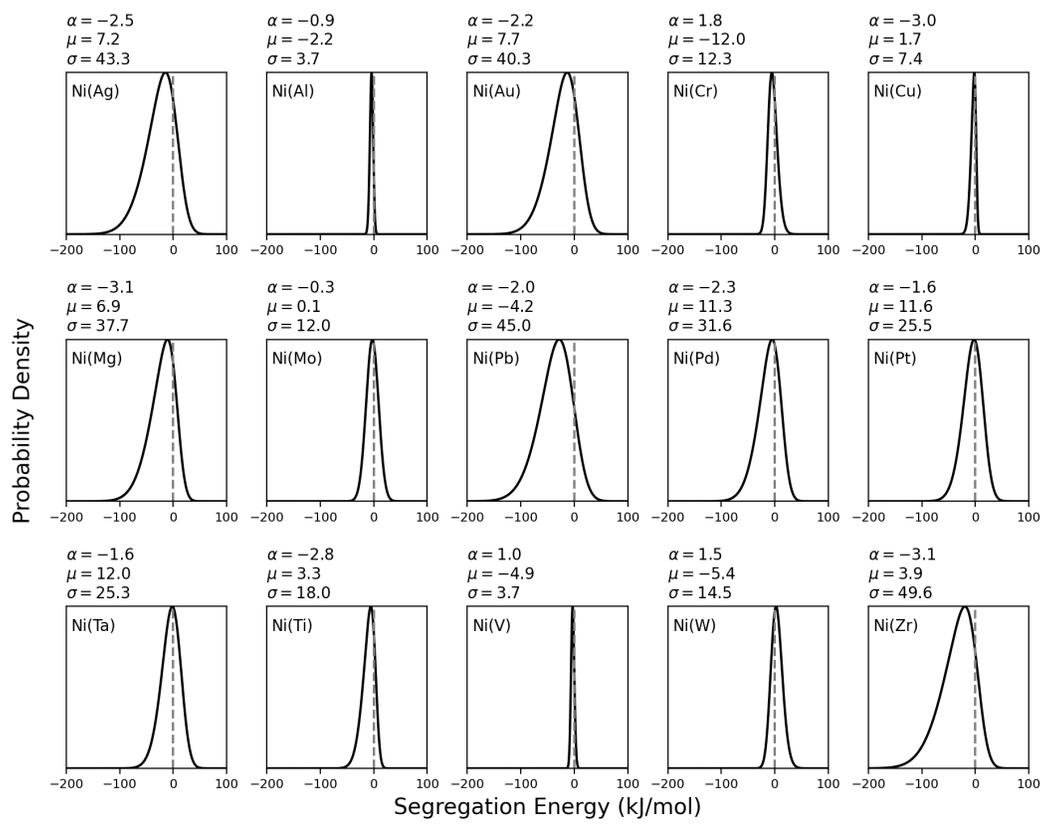

Fig. S8 Ni-based segregation energy spec in kJ/mol.

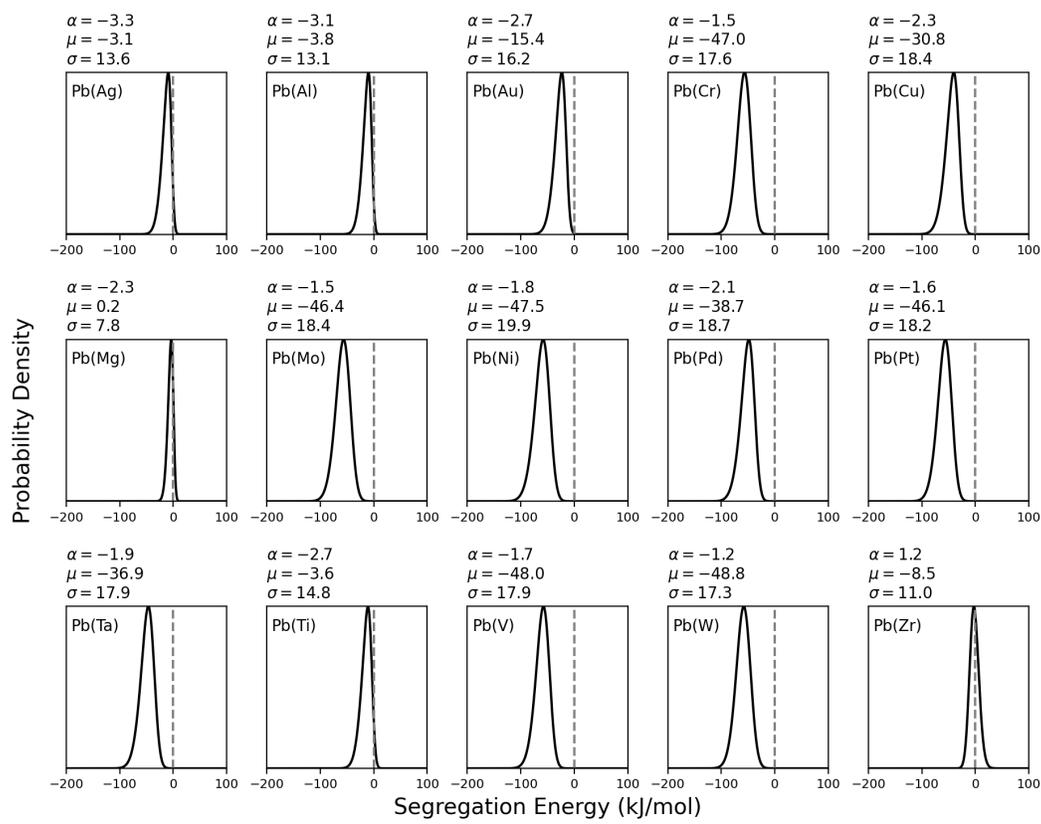

Fig. S9 Pb-based segregation energy spec in kJ/mol.

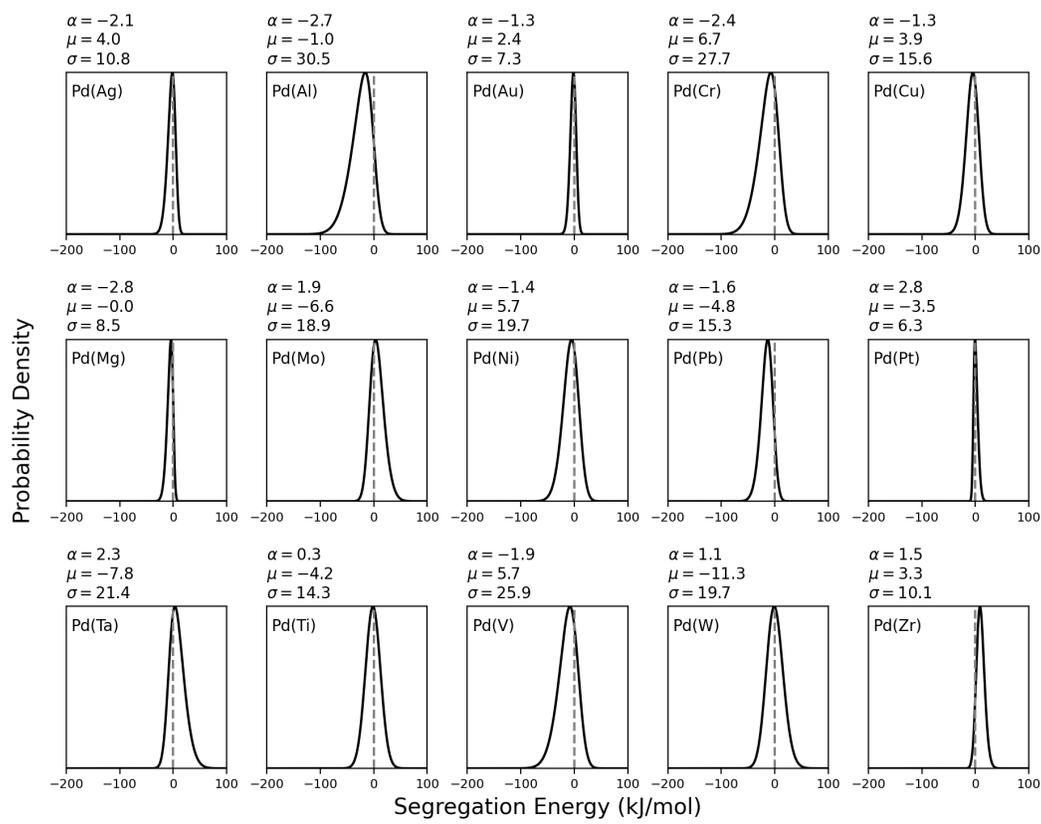

Fig. S10 Pd-based segregation energy spec in kJ/mol.

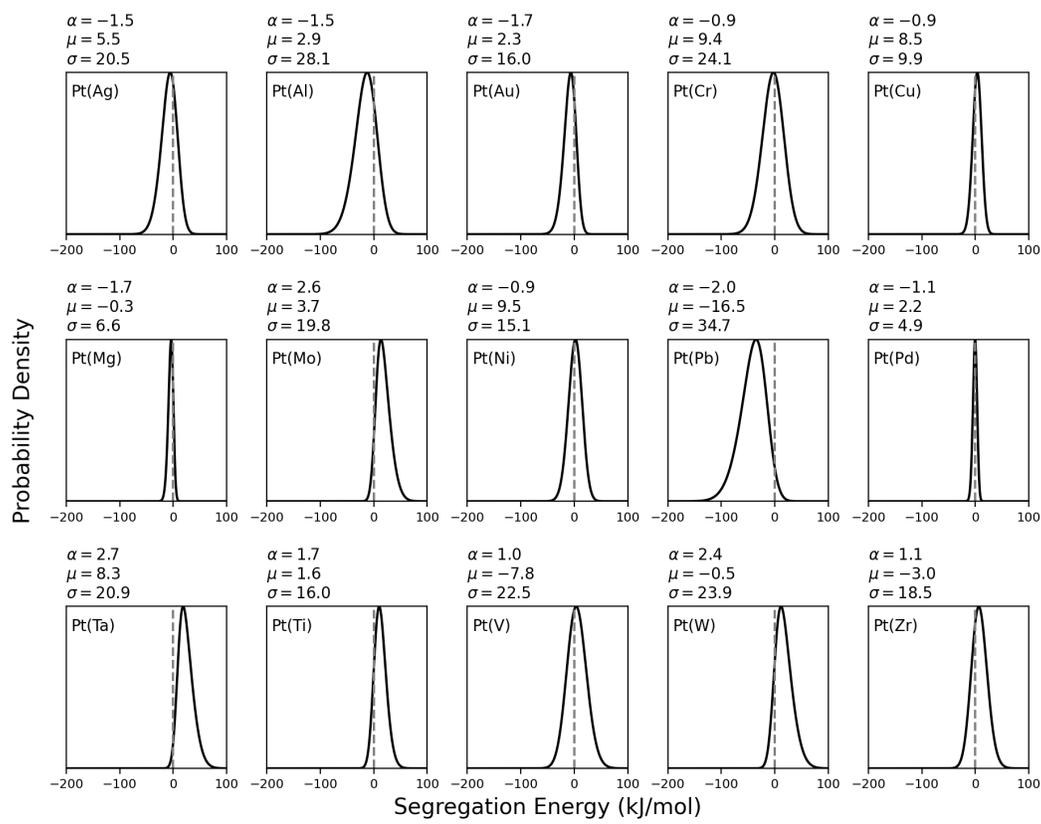

Fig. S11 Pt-based segregation energy spec in kJ/mol.

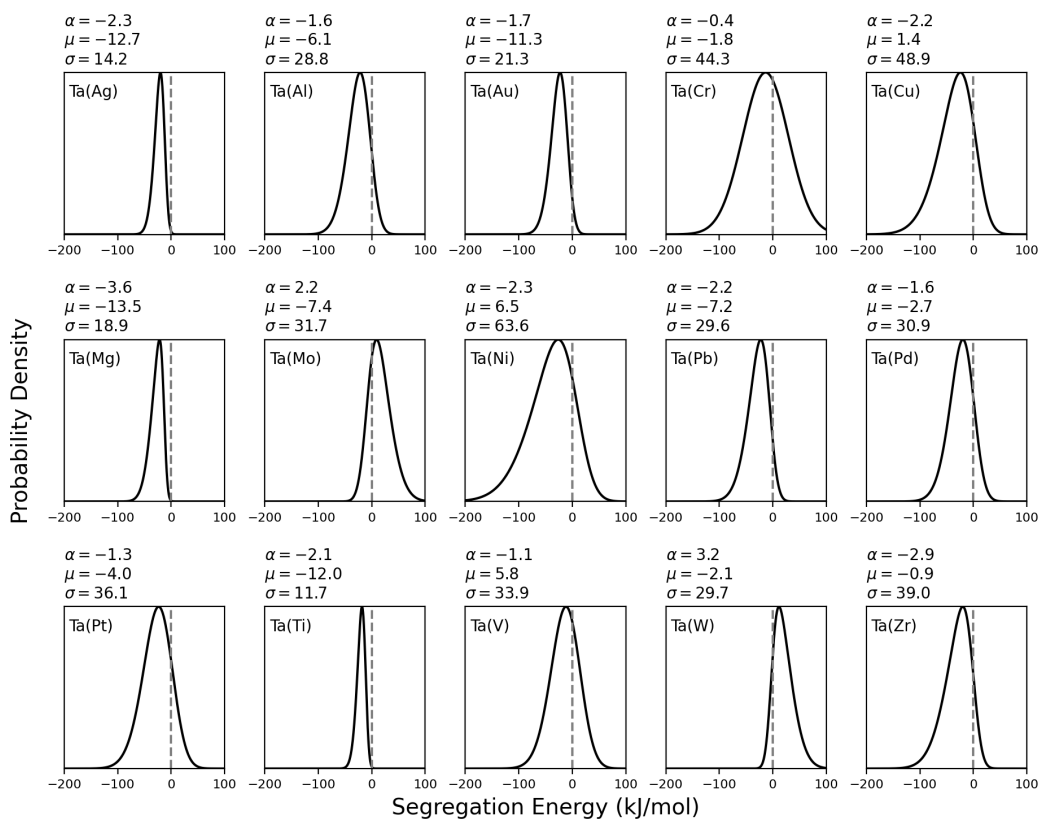

Fig. S12 Ta-based segregation energy spec in kJ/mol.

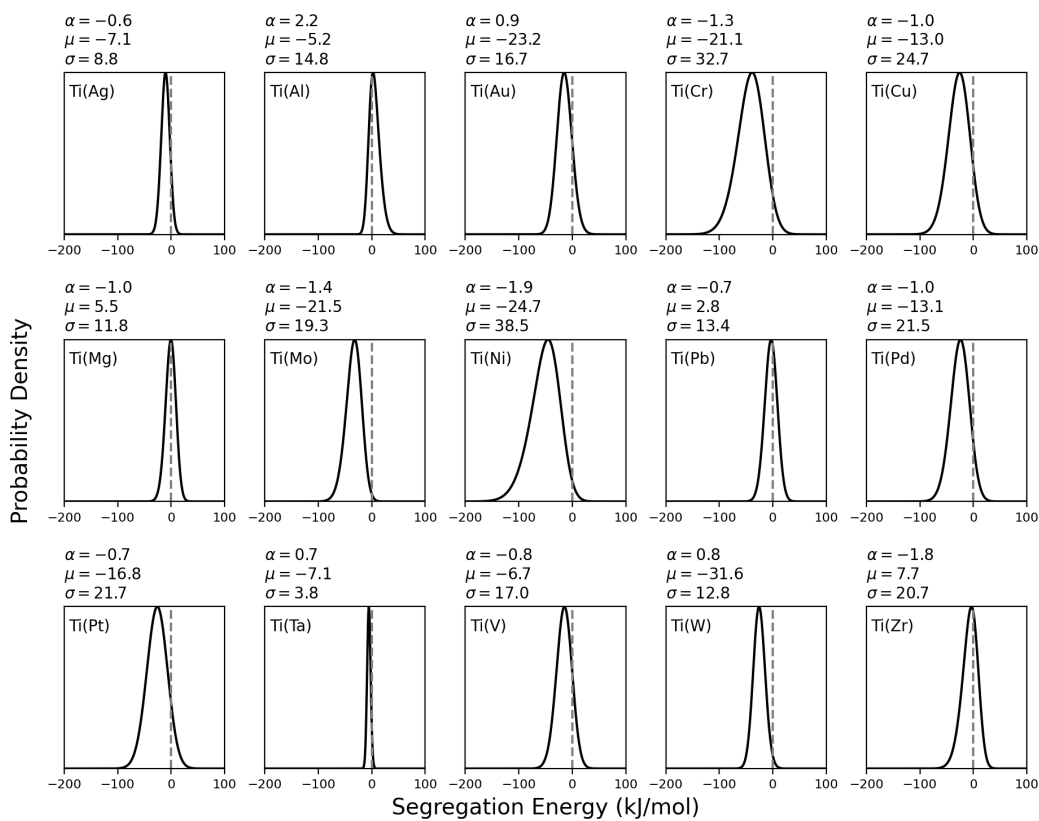

Fig. S13 Ti-based segregation energy spec in kJ/mol.

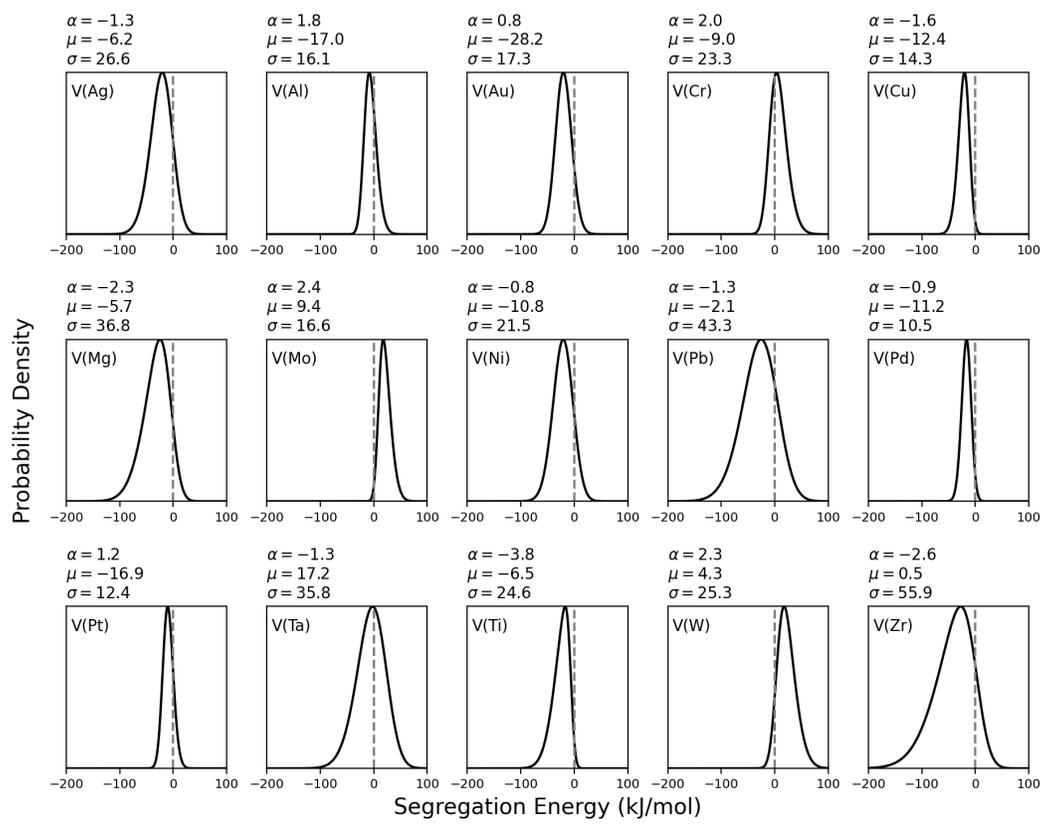

Fig. S14 V-based segregation energy spec in kJ/mol.

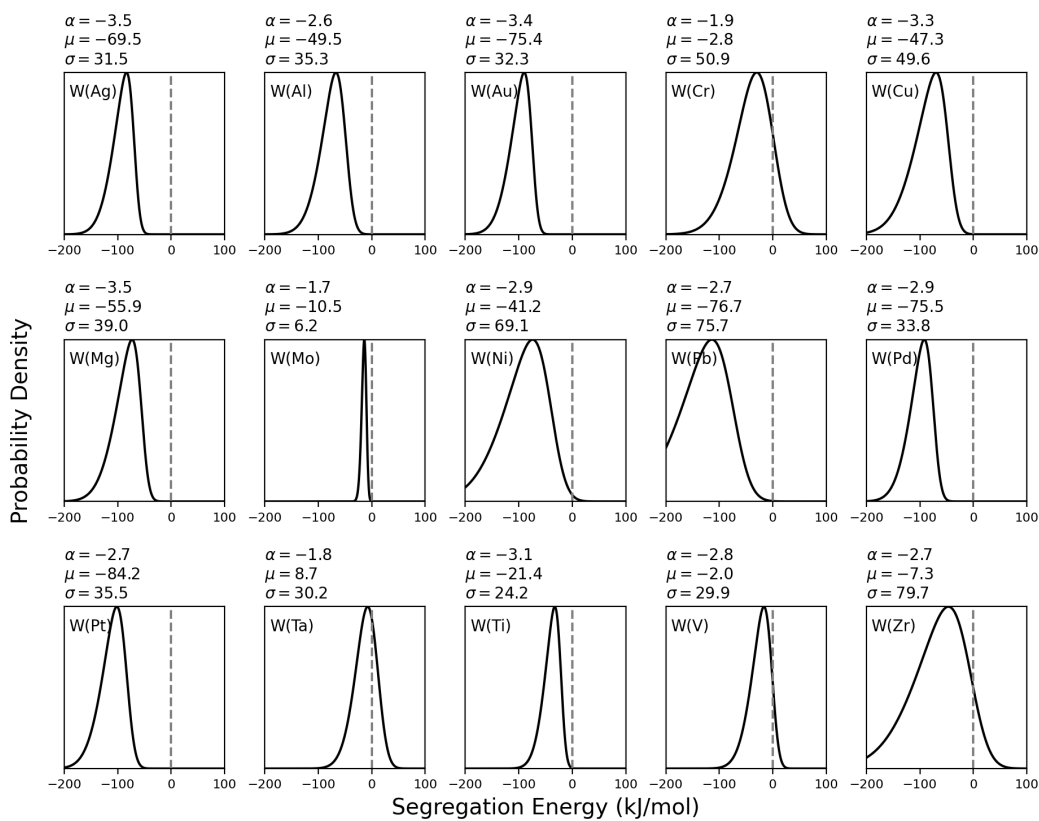

Fig. S15 W-based segregation energy spec in kJ/mol.

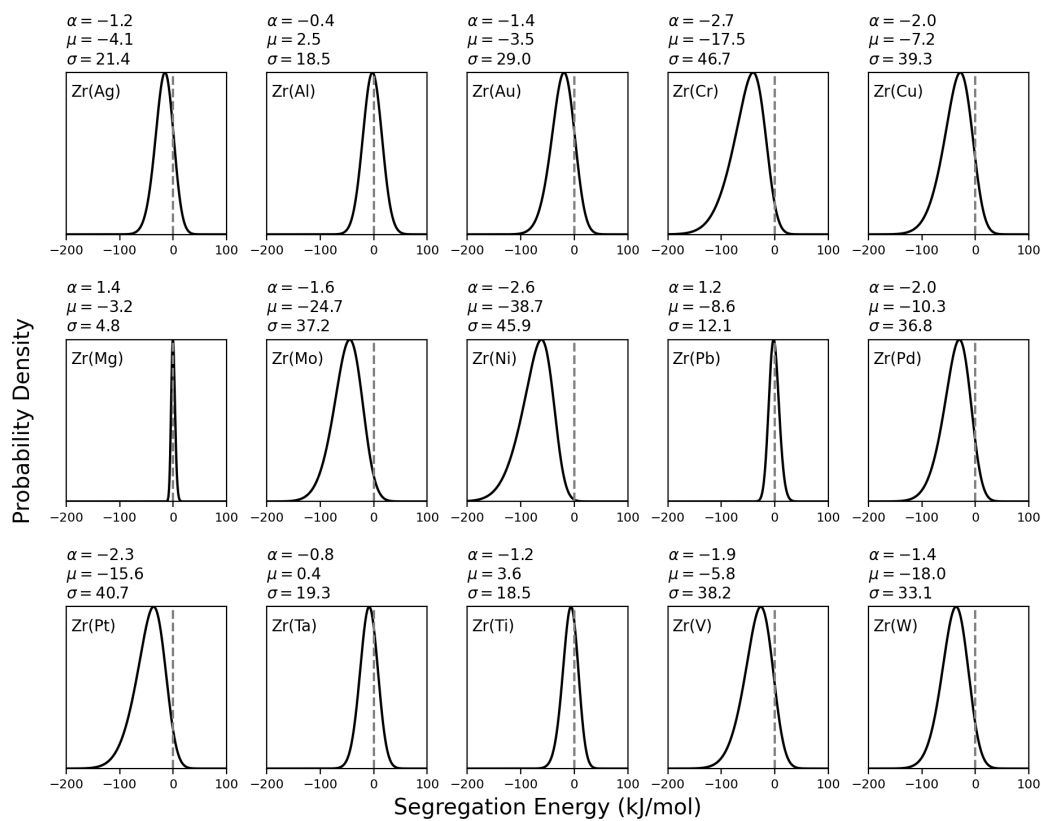

Fig. S16 Zr-based segregation energy spec in kJ/mol.

## 2. Vibrational Segregation Entropy

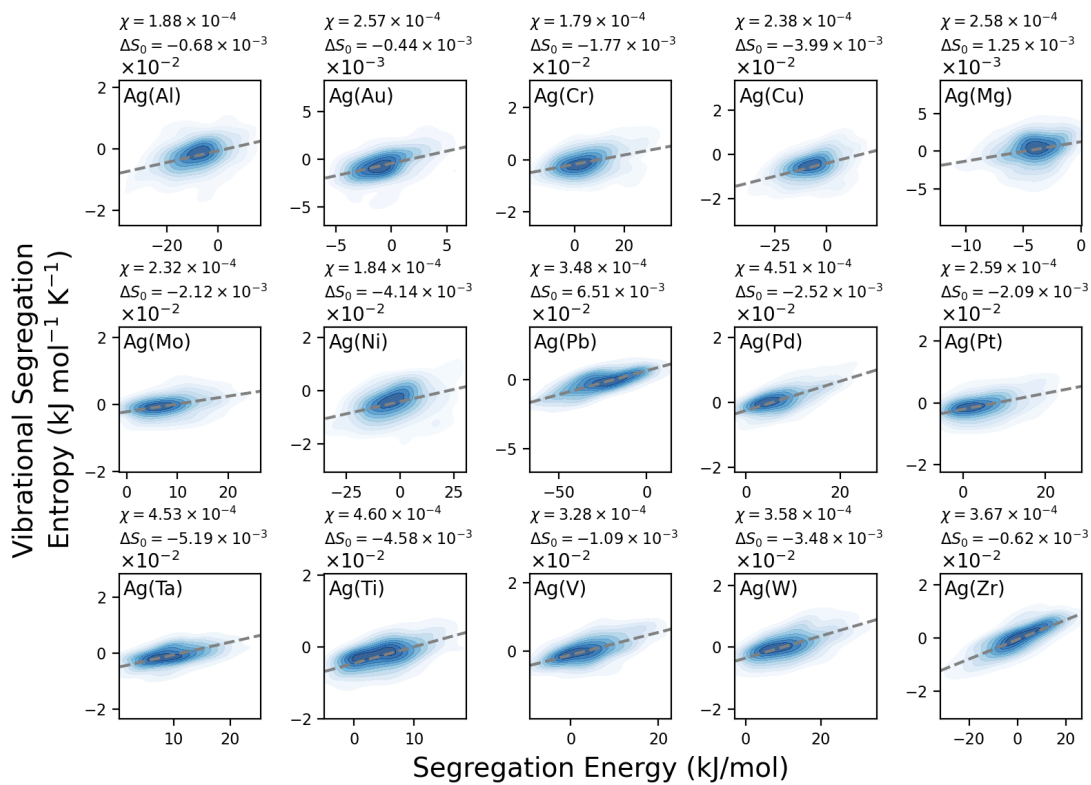

Fig. S17 Ag-based vibrational segregation entropy spectra in kJ·mol$^{-1}$·K$^{-1}$.

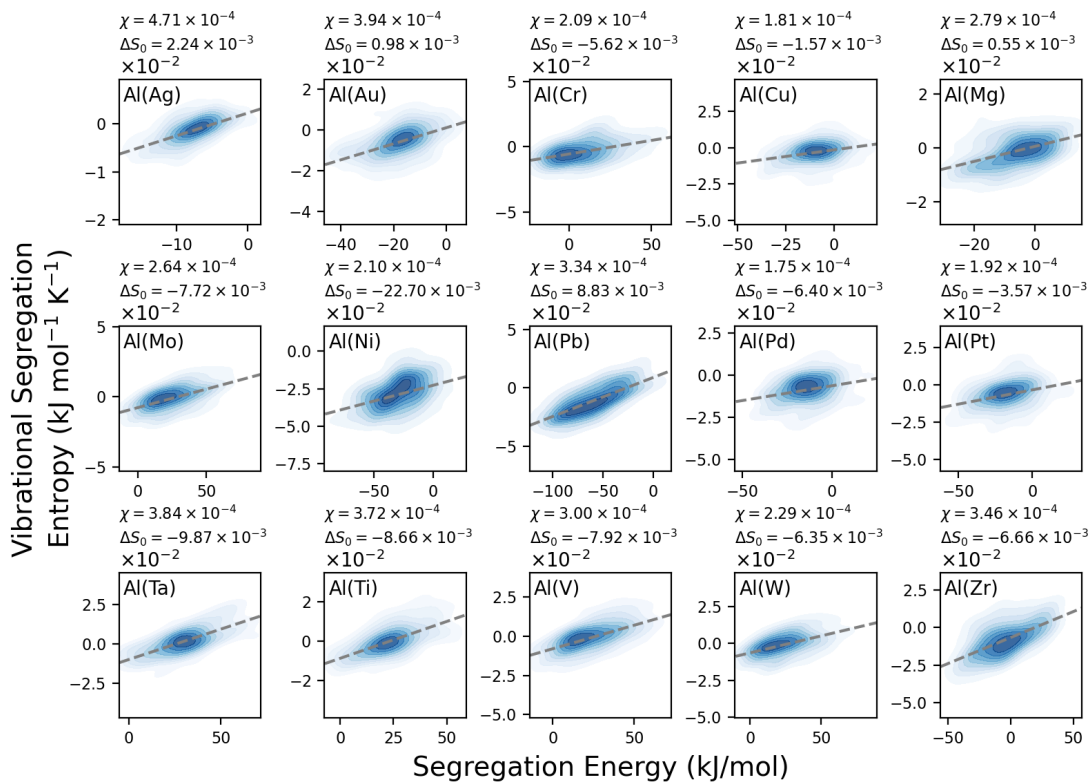

Fig. S18 Al-based vibrational segregation entropy spectra in kJ·mol$^{-1}$·K$^{-1}$.

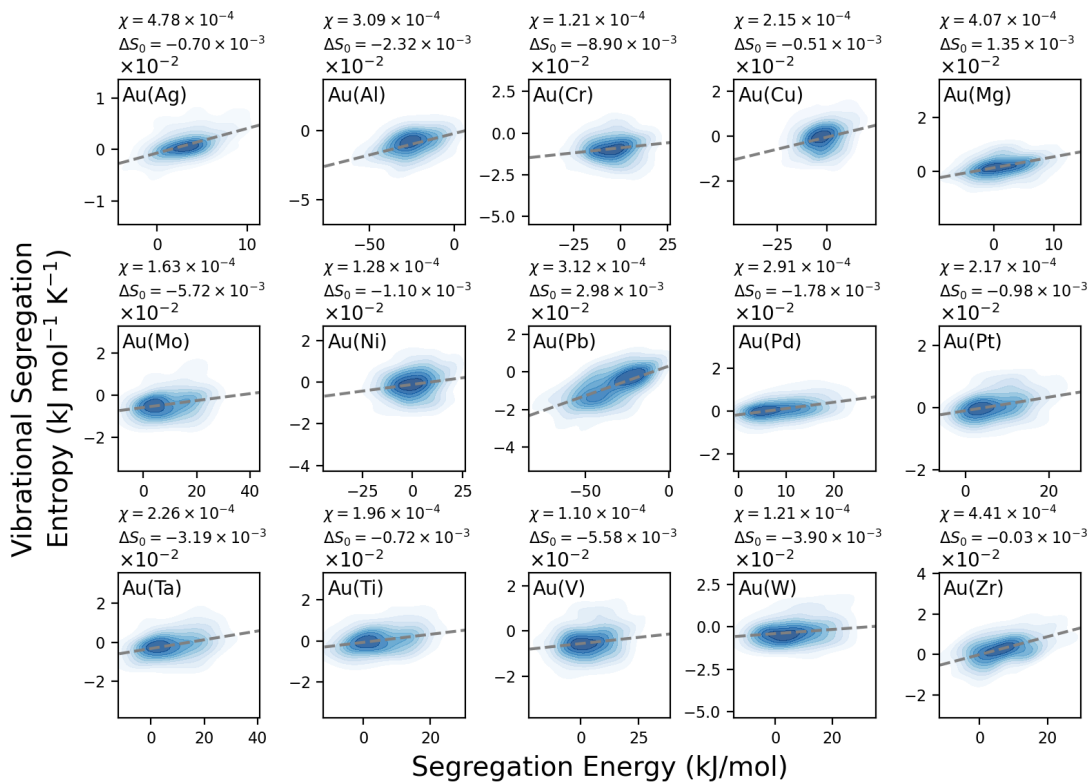

Fig. S19 Au-based vibrational segregation entropy spectra in kJ·mol$^{-1}$·K$^{-1}$.

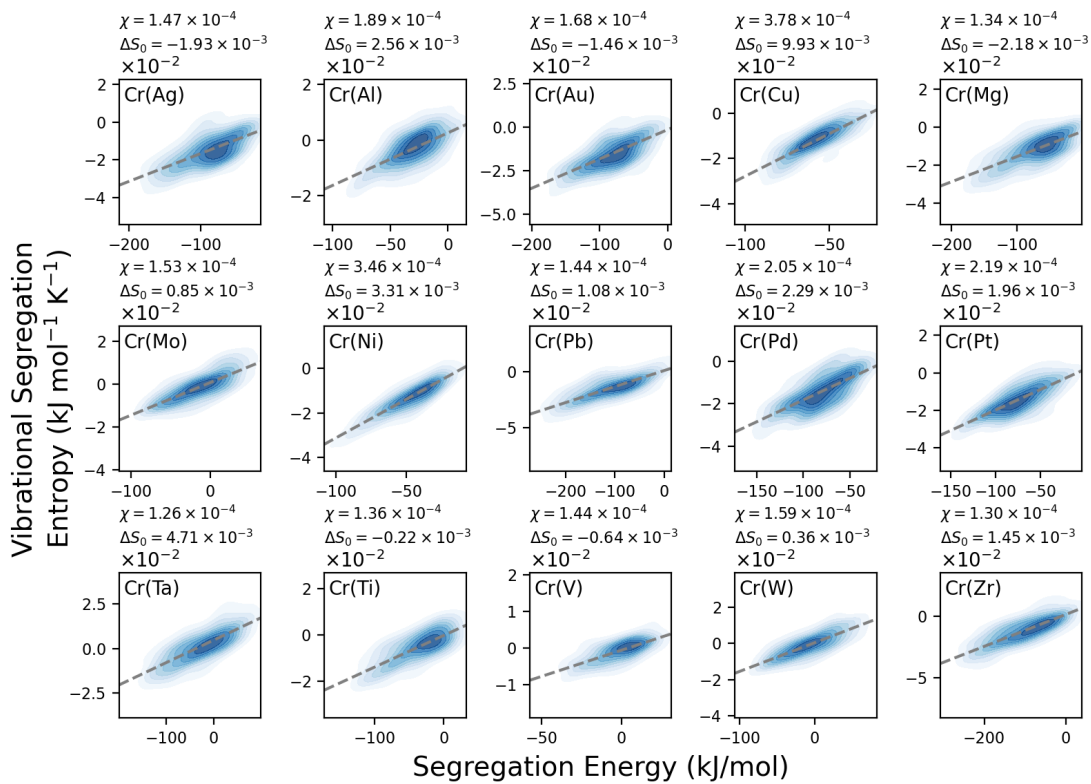

Fig. S20 Cr-based vibrational segregation entropy spectra in kJ·mol$^{-1}$·K$^{-1}$.

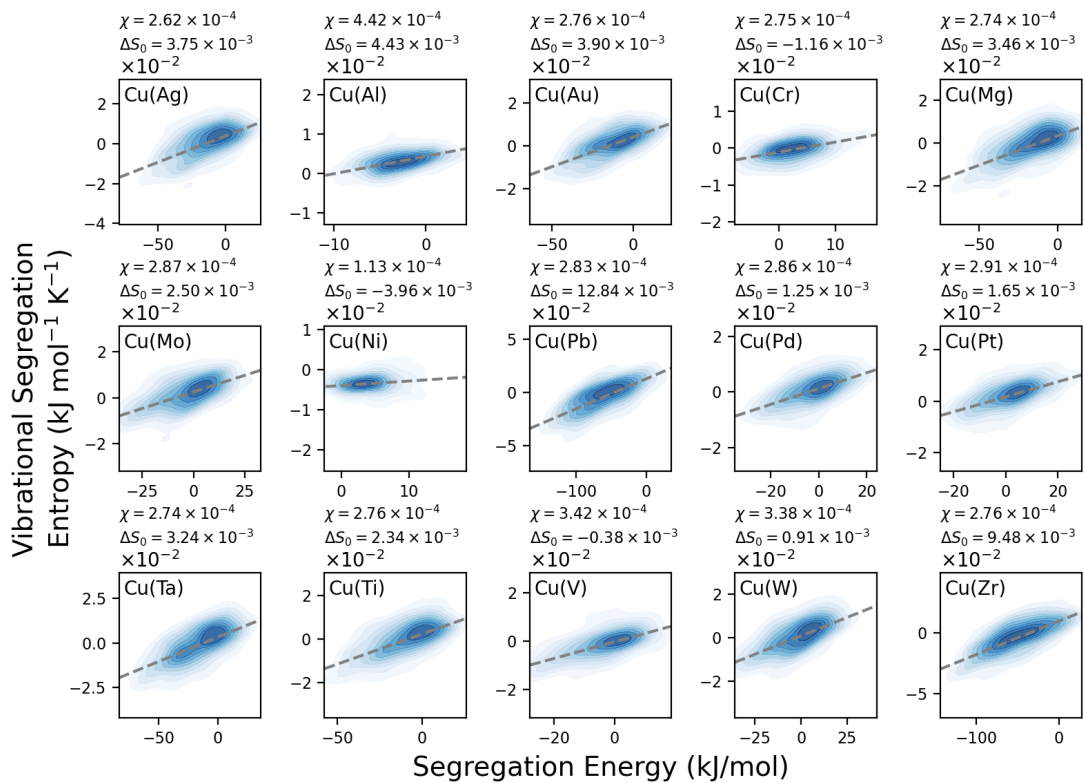

Fig. S21 Cu-based vibrational segregation entropy spectra in kJ·mol$^{-1}$·K$^{-1}$.

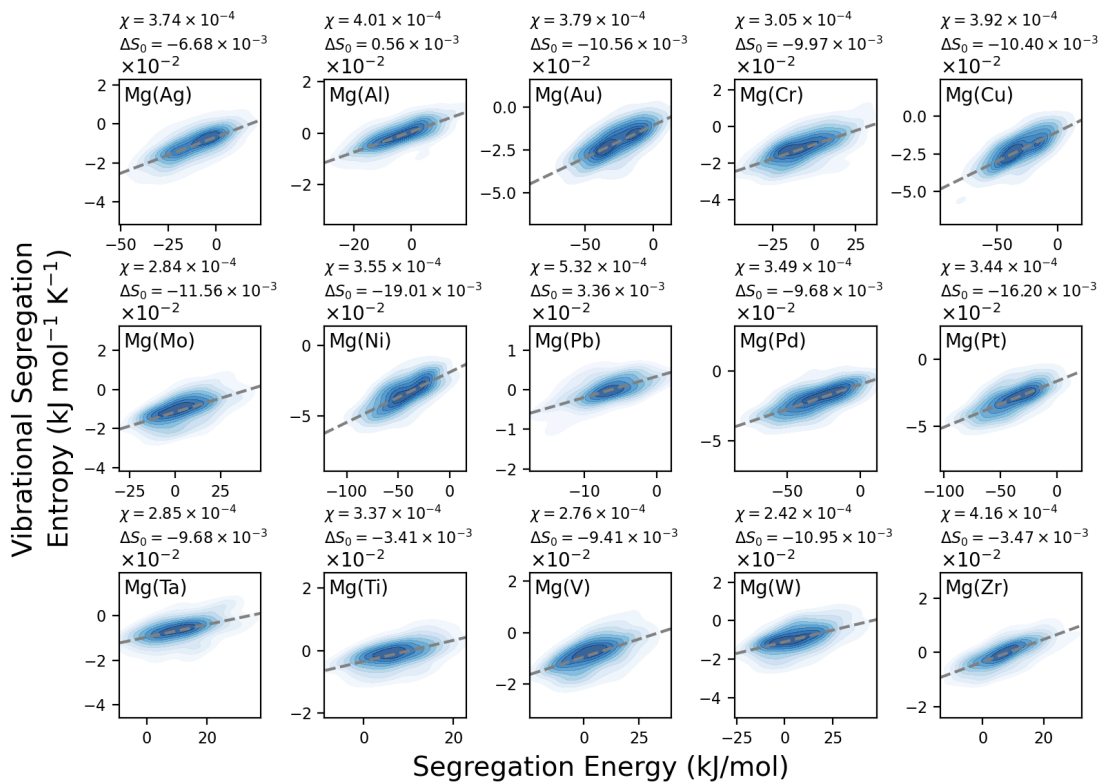

Fig S22 Mg-based vibrational segregation entropy spectra in kJ·mol$^{-1}$·K$^{-1}$.

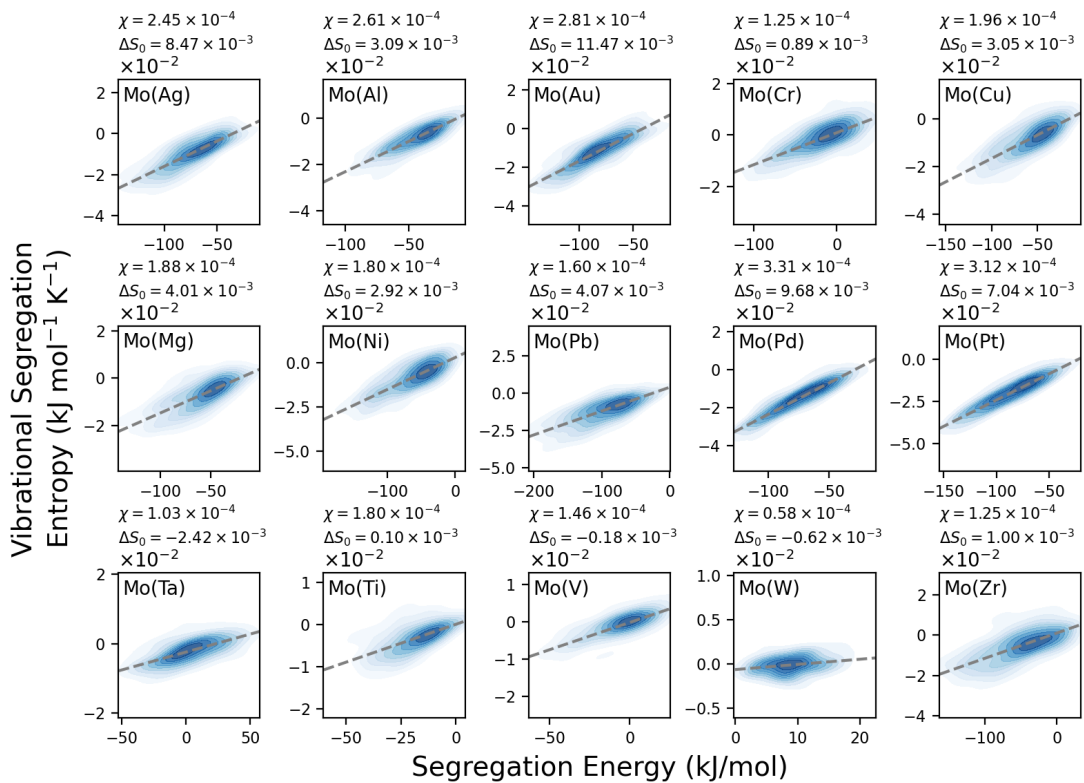

Fig. S23 Mo-based vibrational segregation entropy spectra in kJ·mol$^{-1}$·K$^{-1}$.

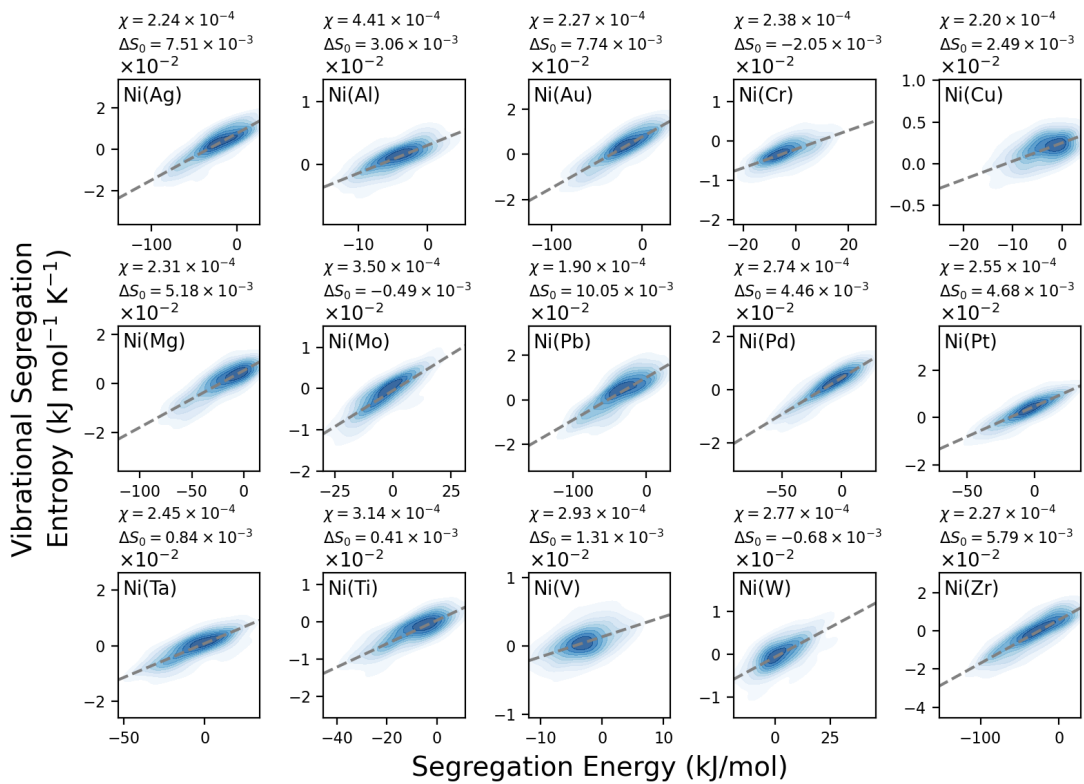

Fig. S24 Ni-based vibrational segregation entropy spectra in kJ·mol$^{-1}$·K$^{-1}$.

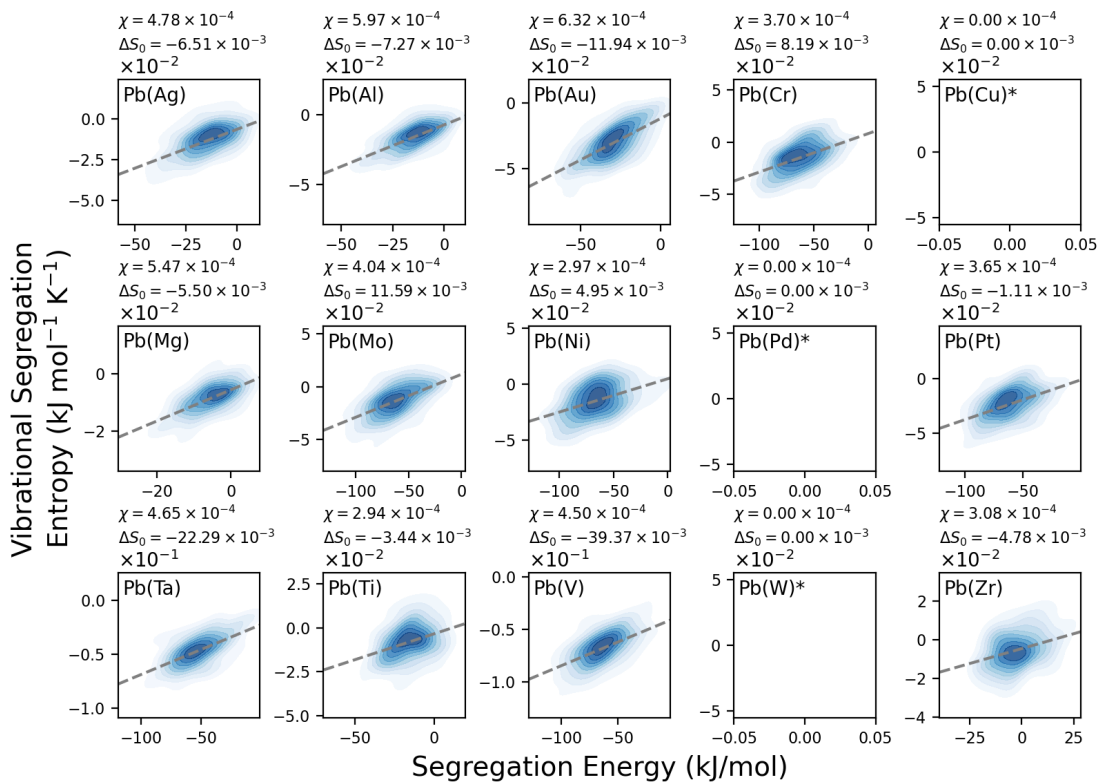

Fig. S25 Pb-based vibrational segregation entropy spectra in kJ·mol$^{-1}$·K$^{-1}$.

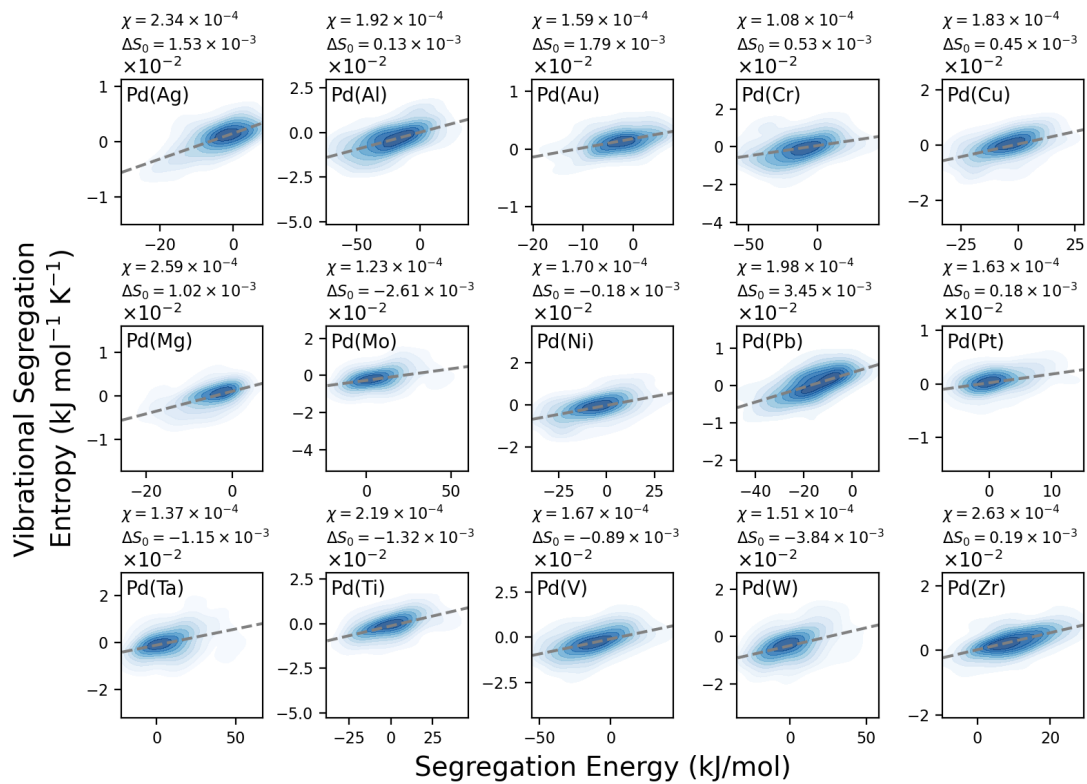

Fig. S26 Pd-based vibrational segregation entropy spectra in kJ·mol$^{-1}$·K$^{-1}$.

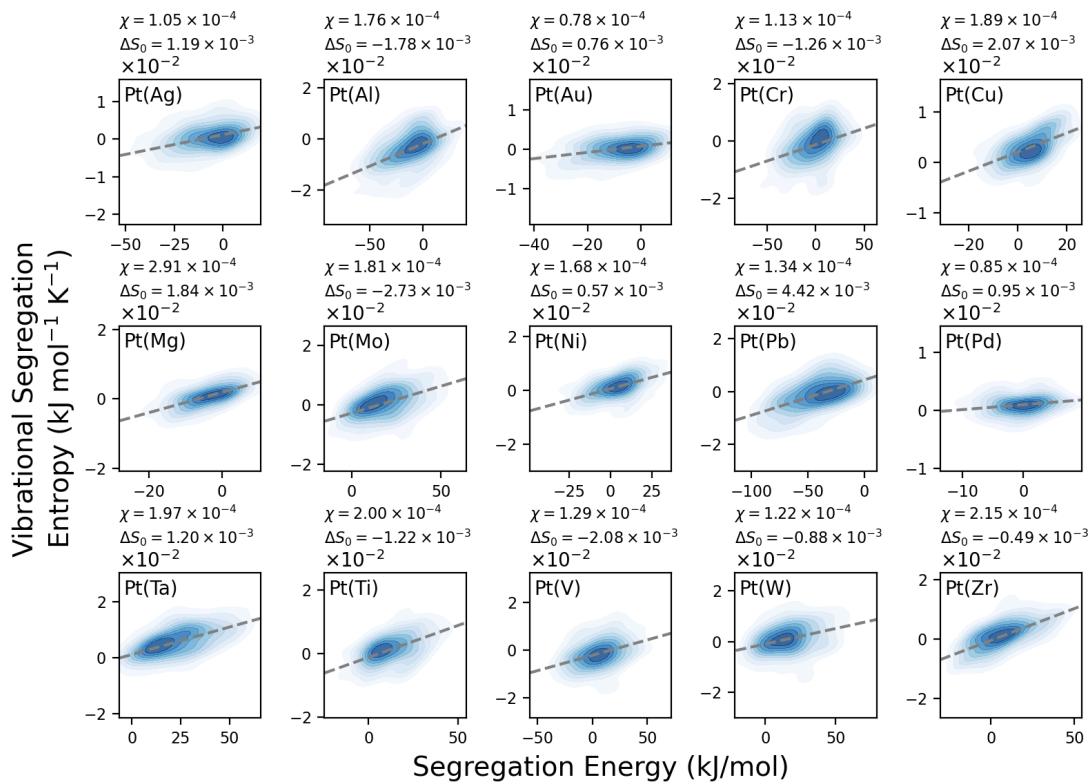

Fig. S27 Pt-based vibrational segregation entropy spectra in kJ·mol$^{-1}$·K$^{-1}$.

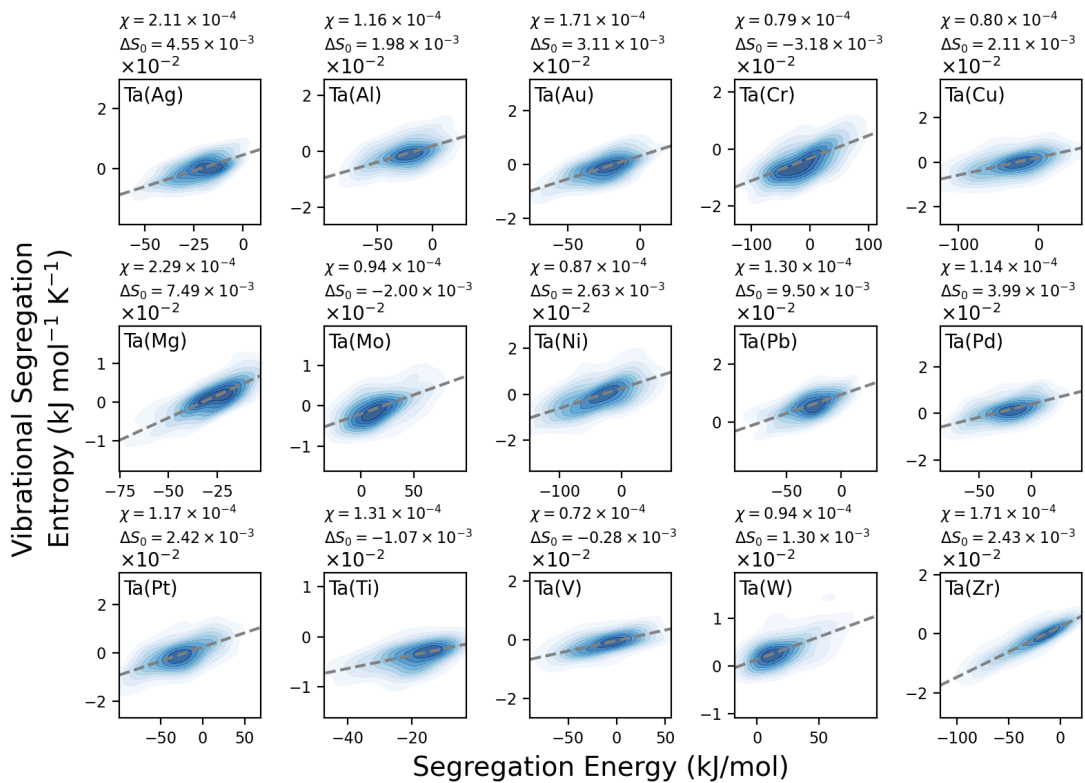

Fig. S28 Ta-based vibrational segregation entropy spectra in kJ·mol$^{-1}$·K$^{-1}$.

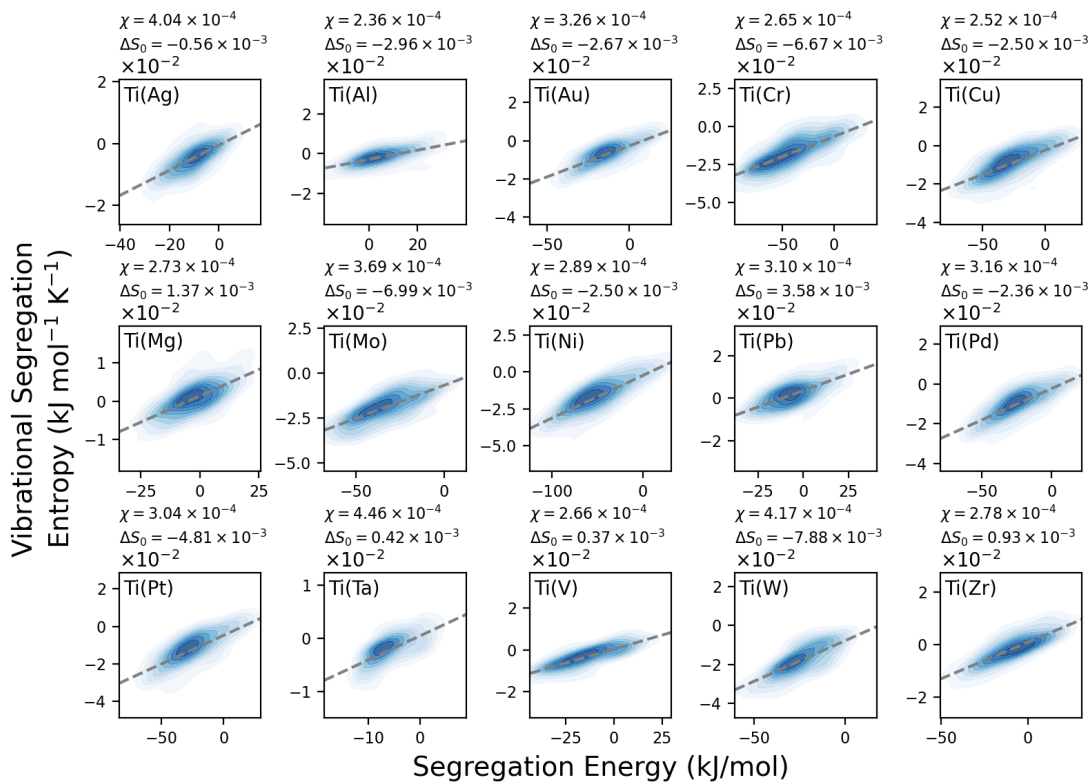

Fig. S29 Ti-based vibrational segregation entropy spectra in kJ·mol$^{-1}$·K$^{-1}$.

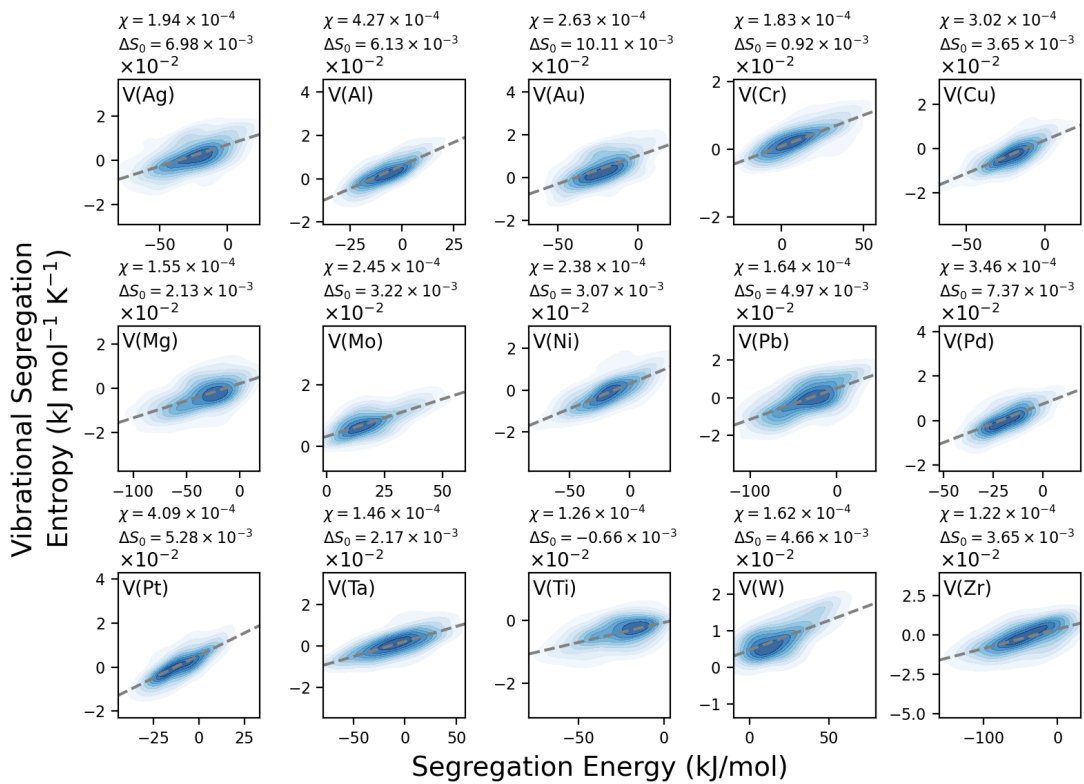

Fig. S30 V-based vibrational segregation entropy spectra in kJ·mol$^{-1}$·K$^{-1}$.

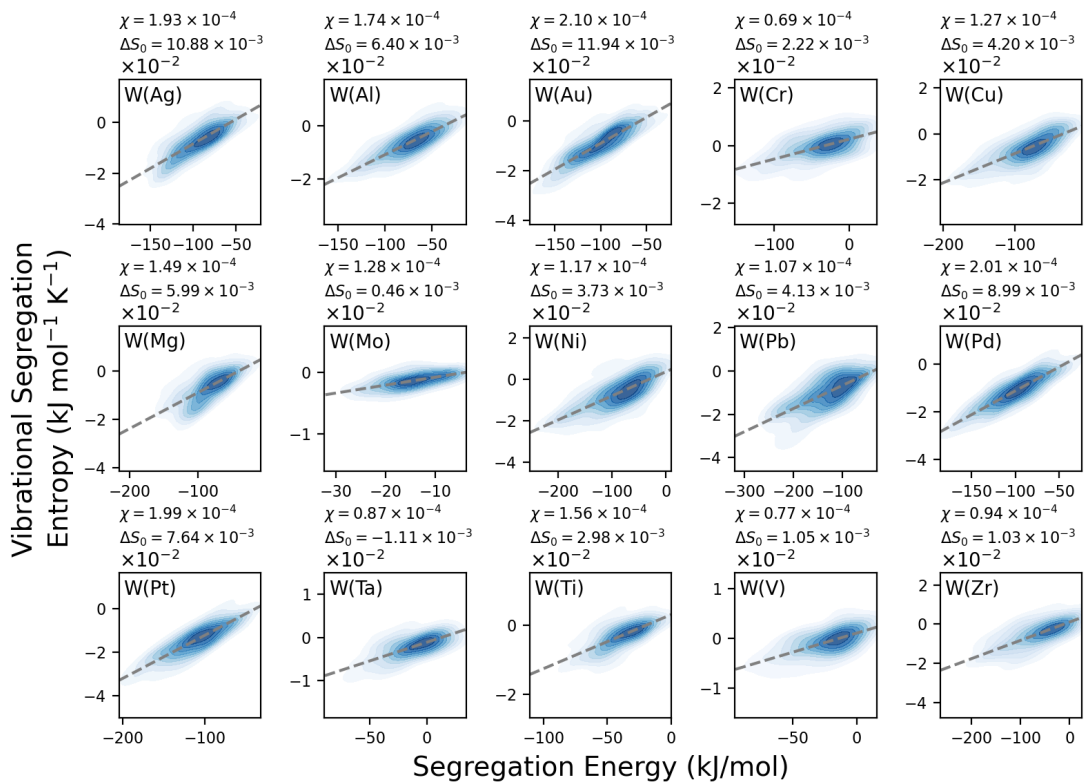

Fig. S31 W-based vibrational segregation entropy spectra in kJ·mol$^{-1}$·K$^{-1}$.

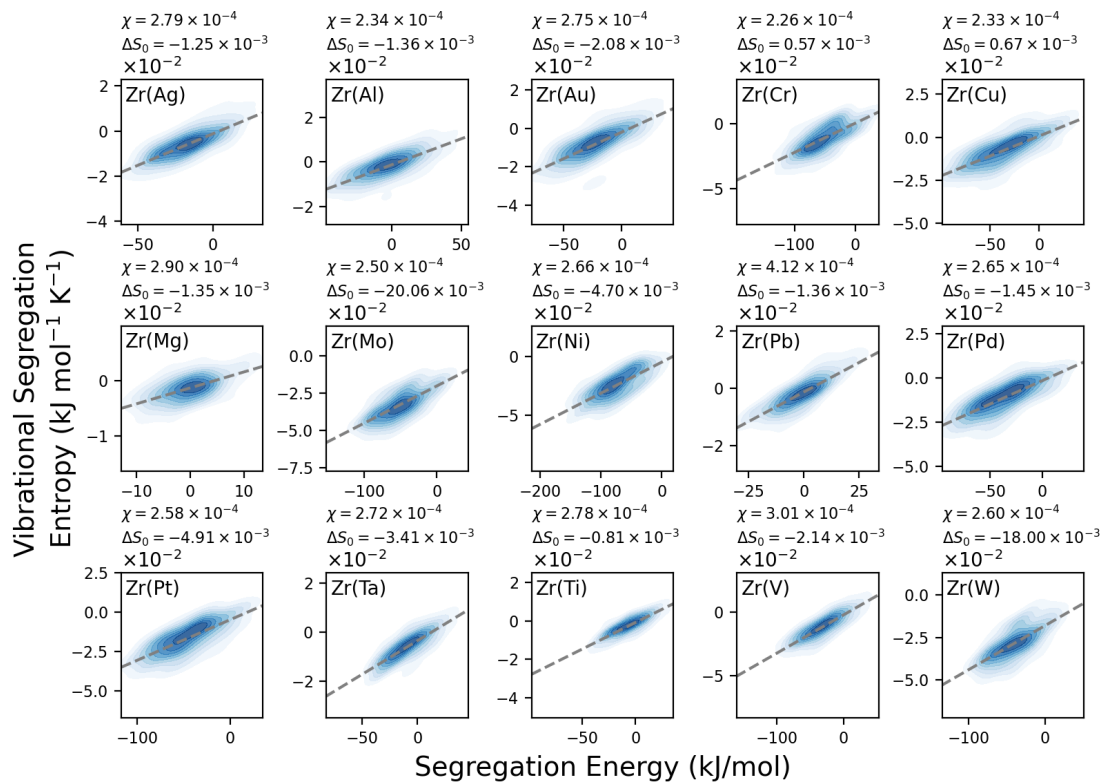

Fig. S32 Zr-based vibrational segregation entropy spectra in kJ·mol$^{-1}$·K$^{-1}$.

## 3. Solute-solute interactions

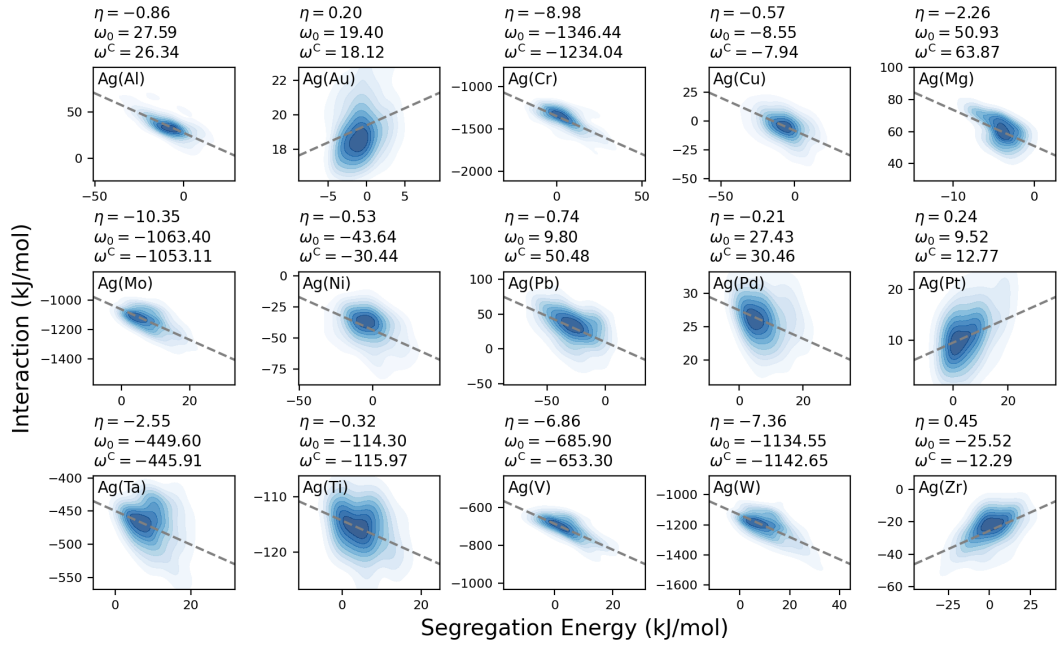

Fig. S33 Ag-based solute-solute interaction spectra in kJ/mol.

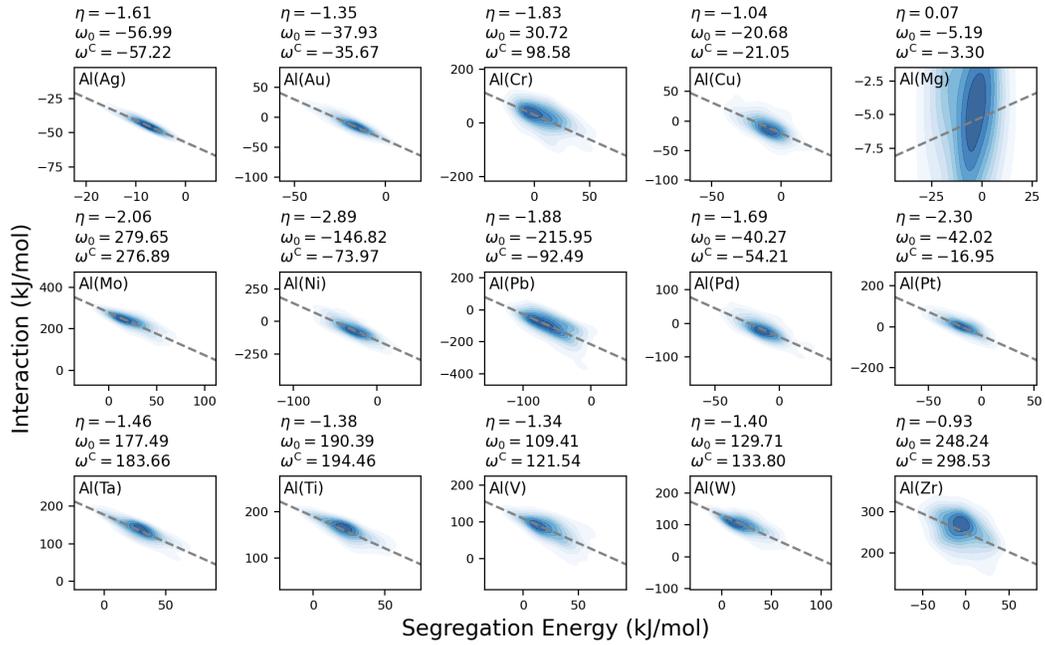

Fig. S34 Al-based solute-solute interaction spectra in kJ/mol.

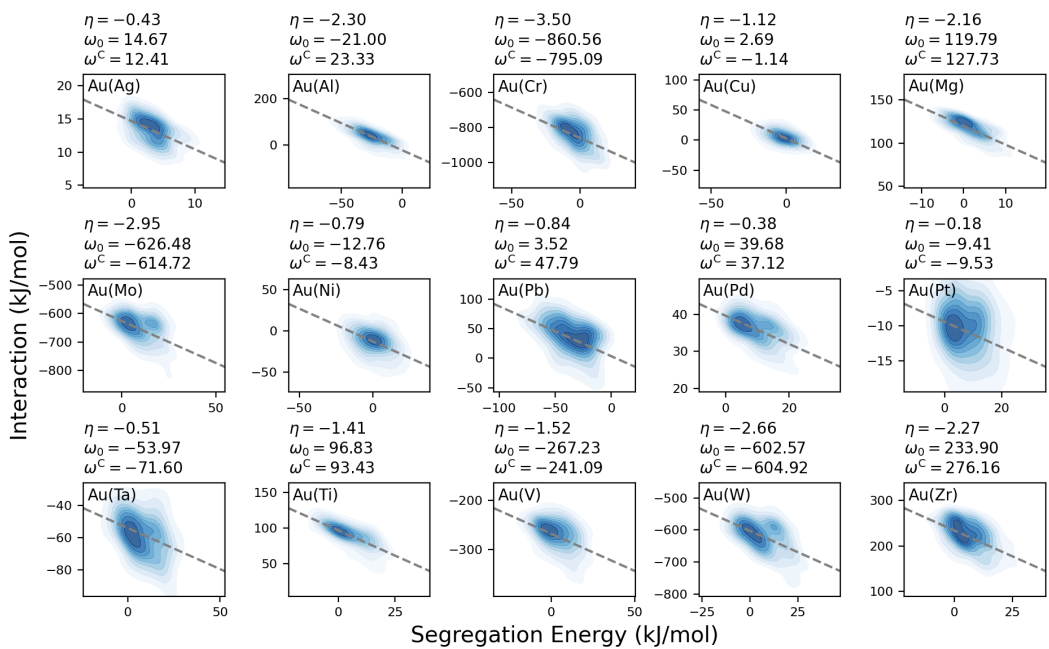

Fig. S35 Au-based solute-solute interaction spectra in kJ/mol.

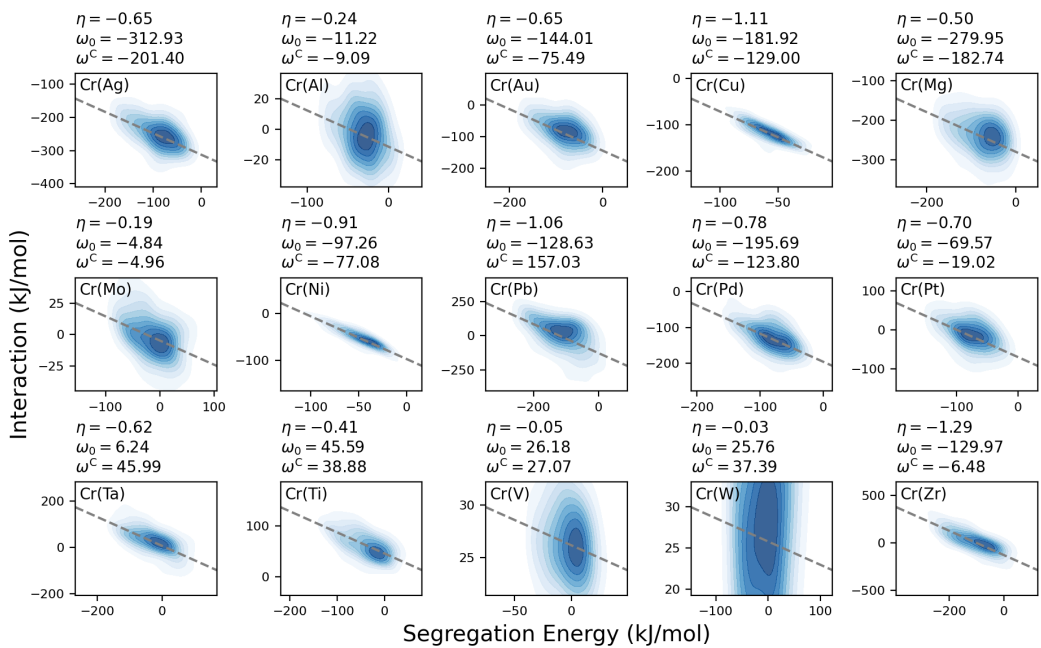

Fig. S36 Cr-based solute-solute interaction spectra in kJ/mol.

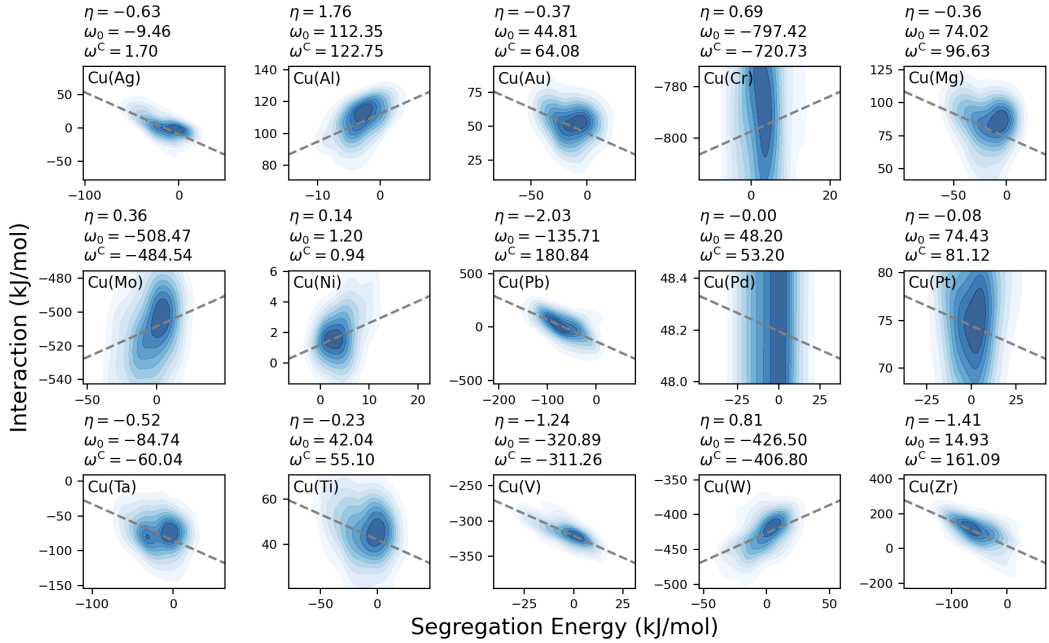

Fig. S37 Cu-based solute-solute interaction spectra in kJ/mol.

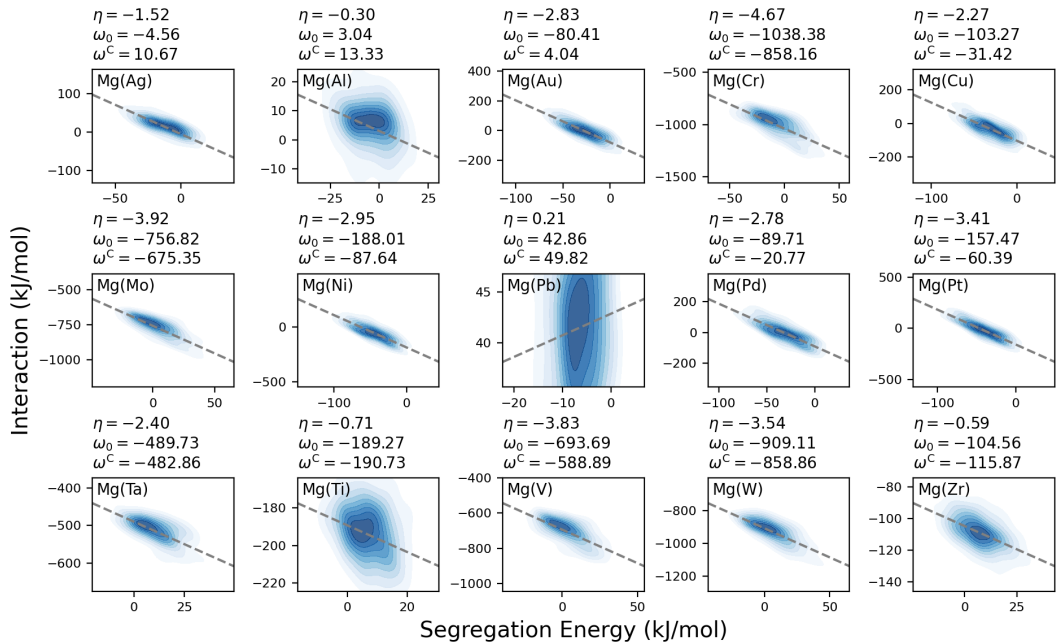

Fig S38 Mg-based solute-solute interaction spectra in kJ/mol.

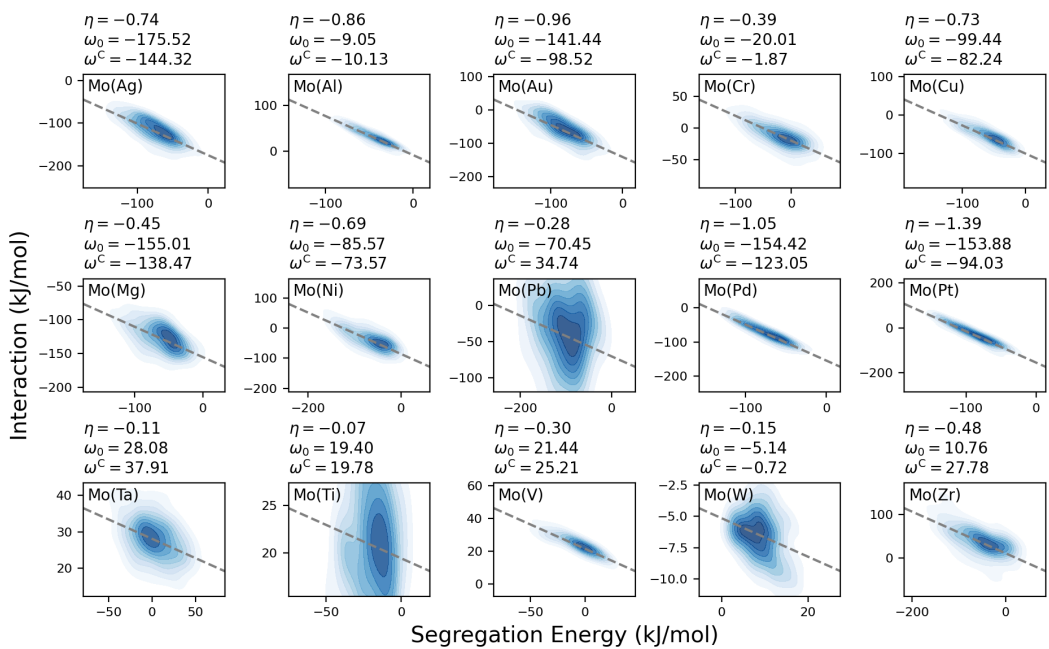

Fig. S39 Mo-based solute-solute interaction spectra in kJ/mol.

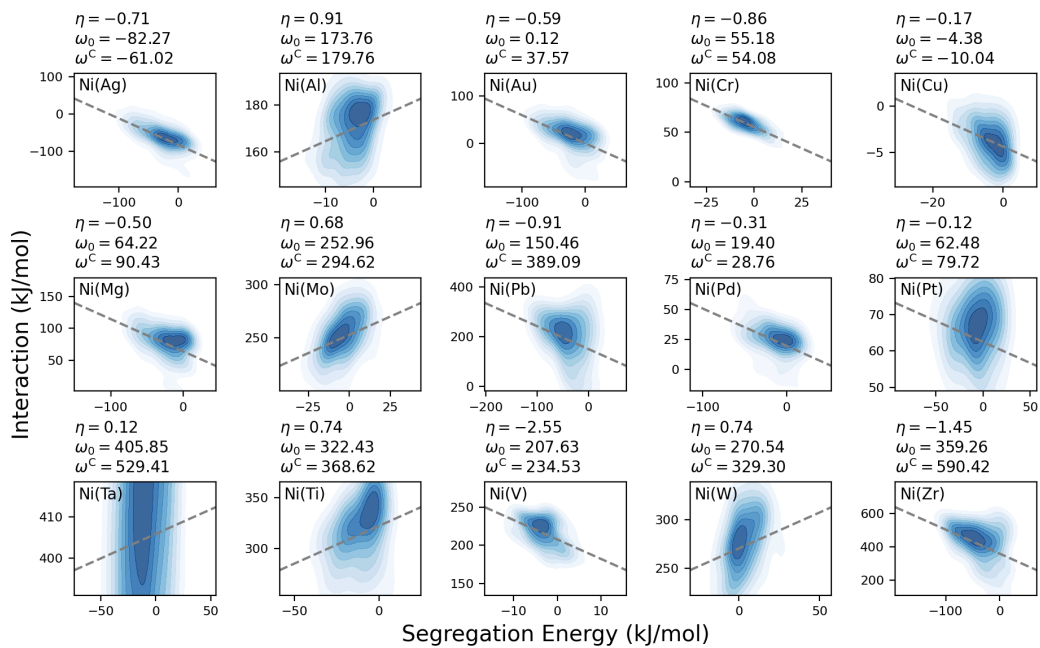

Fig. S40 Ni-based solute-solute interaction spectra in kJ/mol.

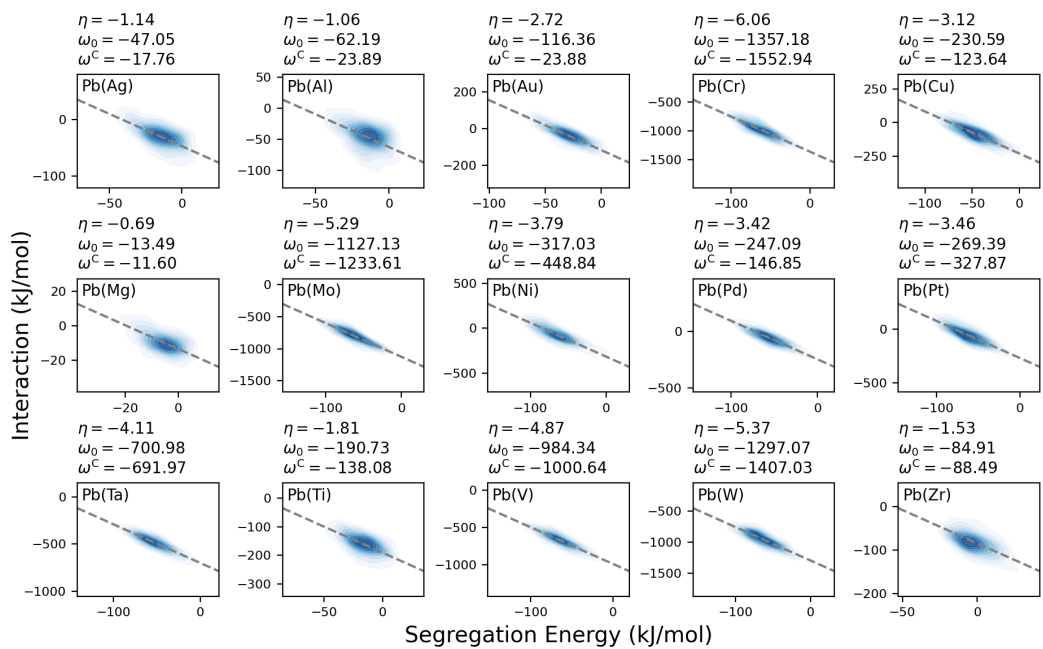

Fig. S41 Pb-based solute-solute interaction spectra in kJ/mol.

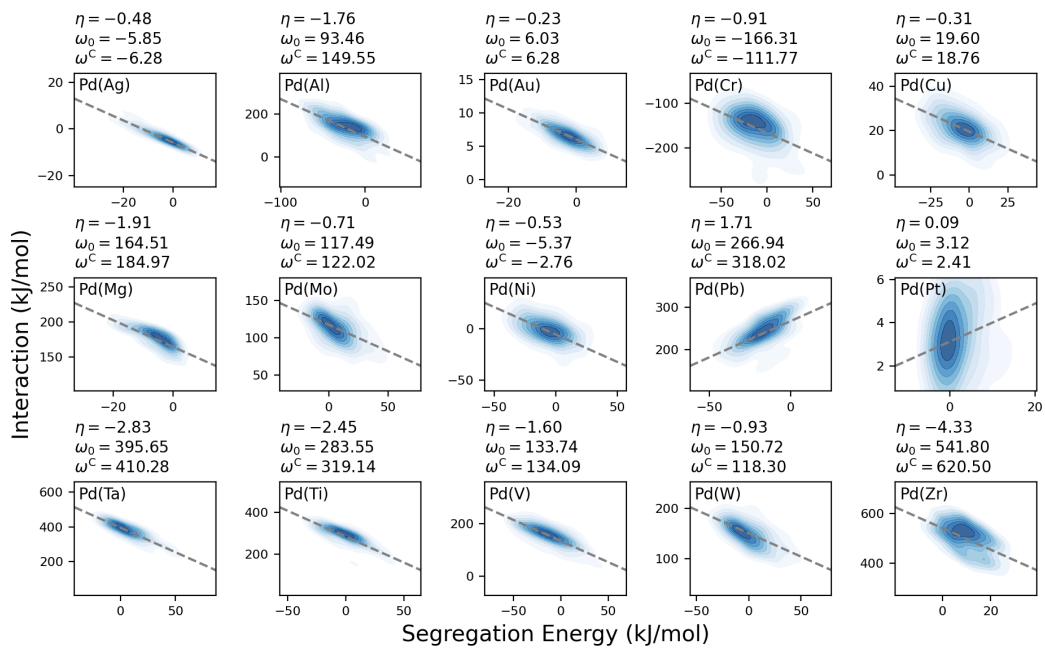

Fig. S42 Pd-based solute-solute interaction spectra in kJ/mol.

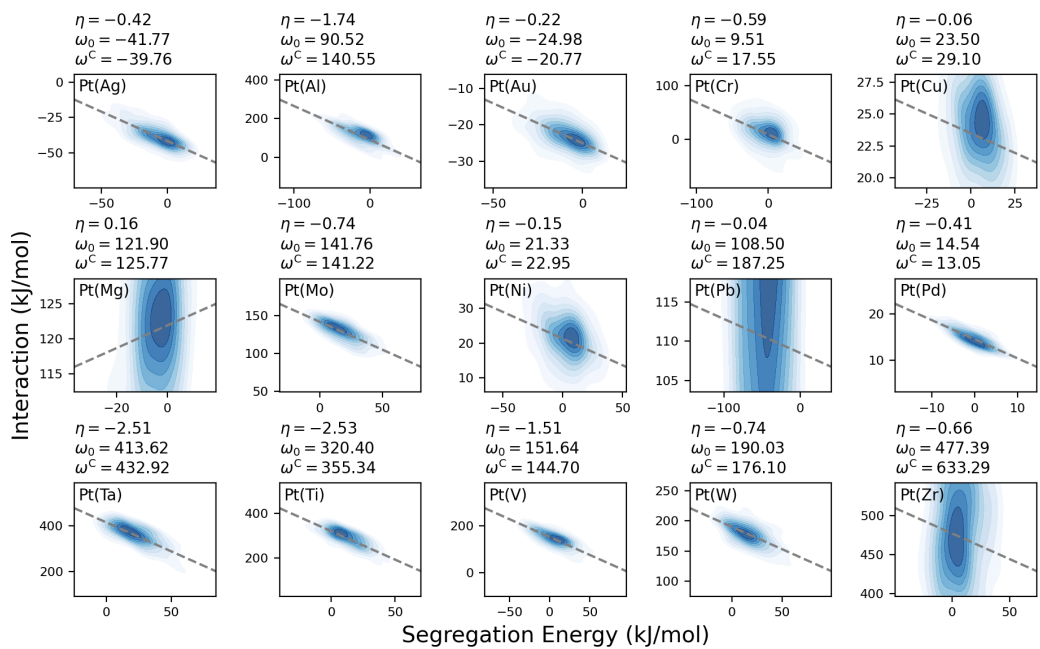

Fig. S43 Pt-based solute-solute interaction spectra in kJ/mol.

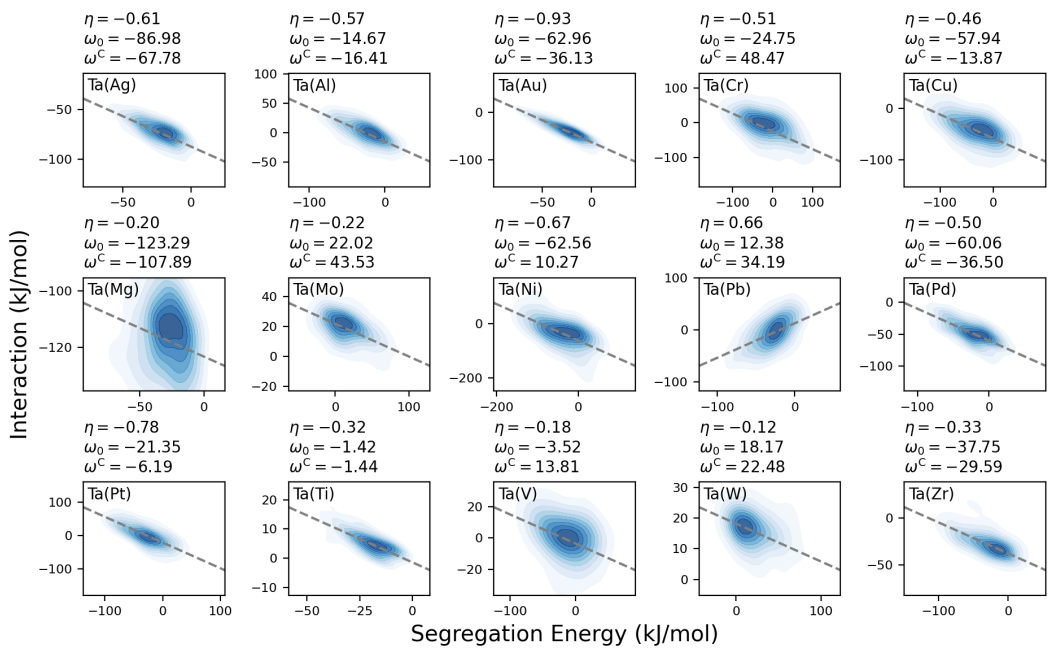

Fig. S44 Ta-based solute-solute interaction spectra in kJ/mol.

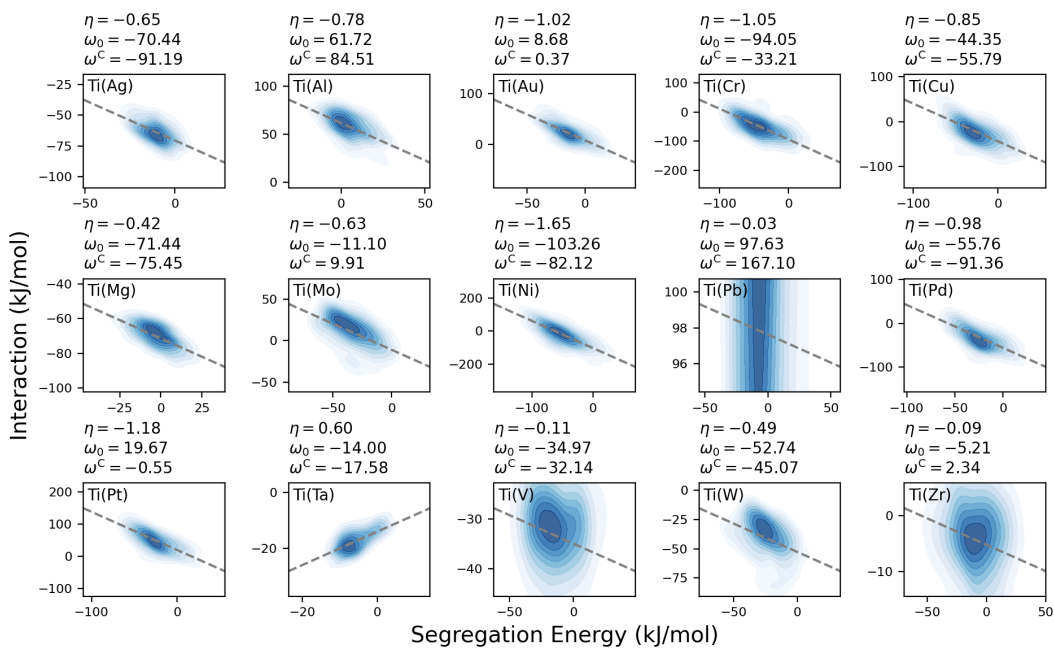

Fig. S45 Ti-based solute-solute interaction spectra in kJ/mol.

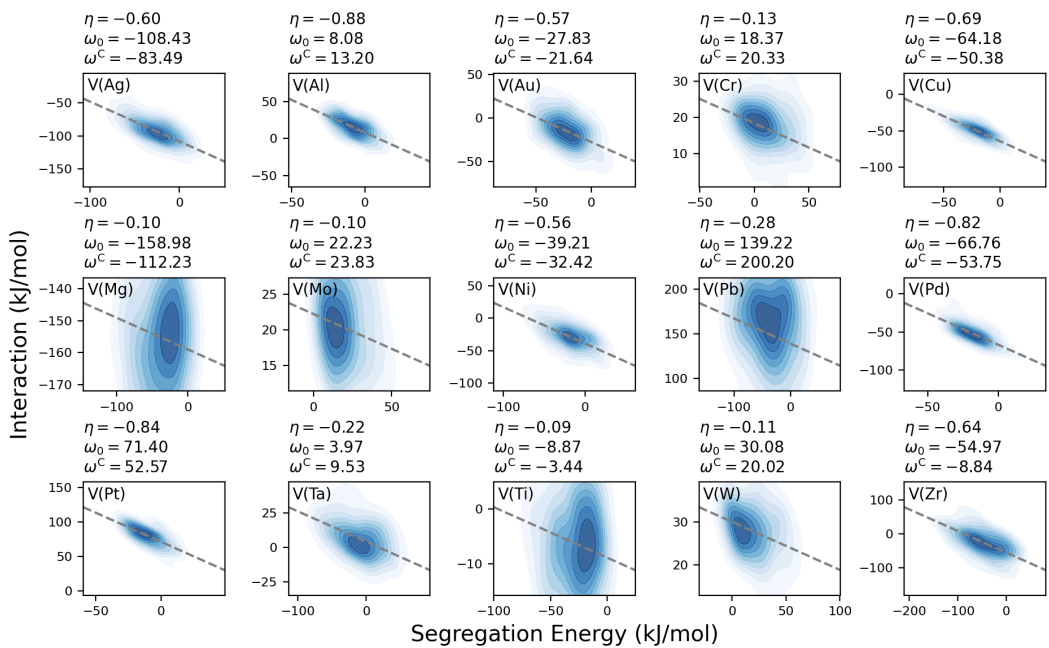

Fig. S46 V-based solute-solute interaction spectra in kJ/mol.

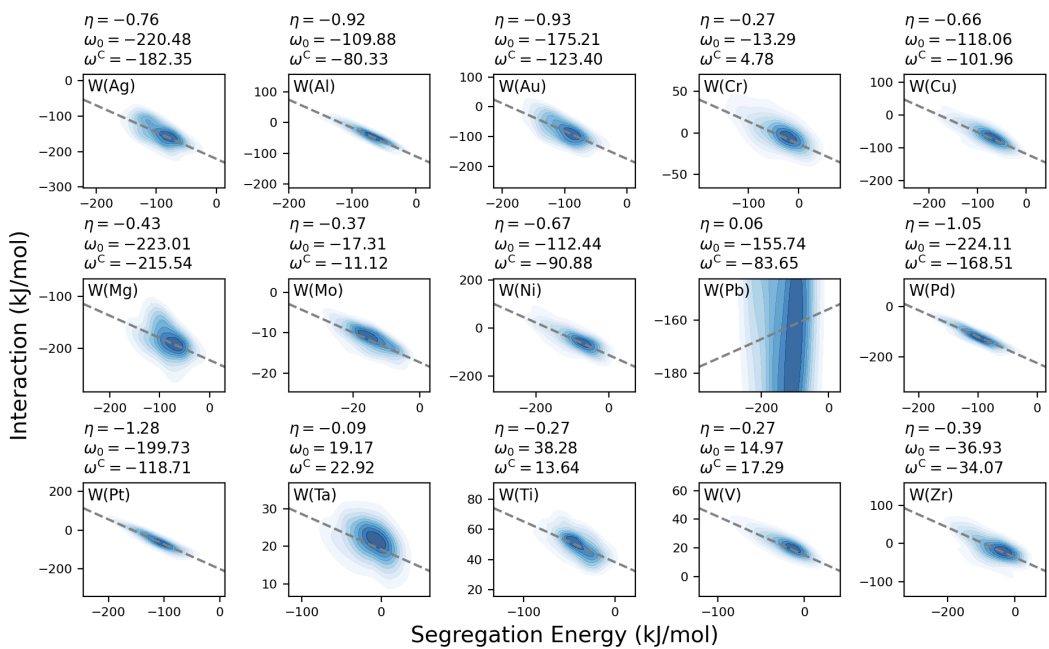

Fig. S47 W-based solute-solute interaction spectra in kJ/mol.

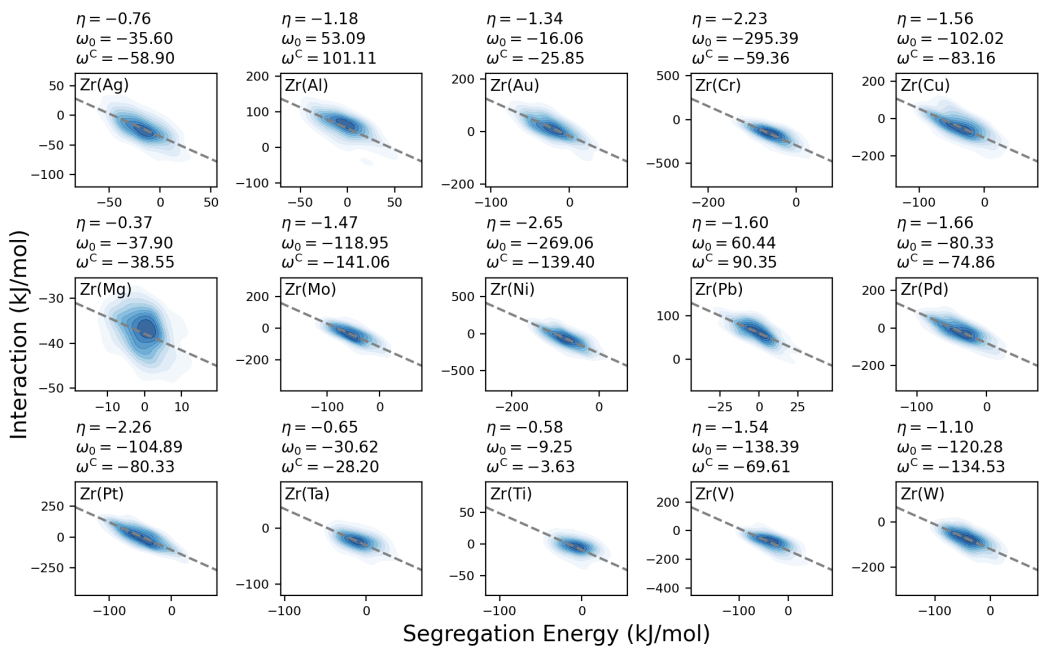

Fig. S48 Zr-based solute-solute interaction spectra in kJ/mol.